Title: Variational band theory of vibronic polarons in crystals. III. Numerical calculations

Authors: M. Georgiev (1,3), A.G. Andreev (2,3), M.D. Ivanovich (3) (Department of Chemistry and Biochemistry (1), University of California San Diego, La Jolla, California and Central Laboratory for Photoprocesses (2) and Institute of Solid State Physics (3) Bulgarian Academy of Sciences, Sofia, Bulgaria)

In the preceding Part II, we derived variational equations for the phonon Fourier amplitudes and for the Fourier amplitudes of the fractional contribution of the electronic bands to the trial variational state. These equations are now solved by means of iterations for each value of the total momentum in order to obtain the energy vs. momentum relation for the ground state. Another result is mapping out the phonon and band Fourier amplitudes in the parameter space of the mixing constant and the electron hopping energy.

## 1. Introduction

In two preceding papers to be referred to as Part I [1] and Part II [2] we surveyed the general premises and derived variational equations, respectively, for describing vibronic polarons itinerant along linear atomic chains. The variational eigenstate was a linear combination of trial states for either of the electronic bands. It should be stressed that in our model the constituent electronic bands coupled to the vibration through the electron-phonon mixing term, band-off-diagonal unlike the commonly used band-diagonal coupling terms. This distinguished our physics from the traditional Holstein problem. Our method of extending Merrifield's Variational Ansatz [3.4] to a system of more than one electronic band was transparent physically, though it might not be free of deficiencies.

Apart from its purely academic interest, our variational study as described in Part II applies directly to a few specific physical systems under extensive research for some time, such as the high-temperature superconducting cuprates [5] and colossal magnetoresistance exhibiting manganates [6]. Indeed the physical features of both materials seem largely dependent on the charge transfer along metal-oxide chains. In general, the variational method of diagonalizing the vibronic Hamiltonian [7] is expected to generate either Jahn-Teller polarons if the progenitor electronic bands are degenerate or Pseudo-Jahn-Teller polarons if these bands are nearly degenerate [8]. Both species are considered by many to be the likely charge carriers in copper- and manganese- oxygen manifolds.[5]

## 2. Numerical calculations

The $\beta$-variational equations as obtained from $\partial/\partial\beta = 0$ read [2]:

$$\beta_{q\mu}^{\kappa} = \{(g^{\mu\mu}/\eta\omega) + (g^{\mu\nu}/\eta\omega)(\alpha_\nu^{\kappa}/\alpha_\mu^{\kappa})S_{0\kappa}^{\nu\mu}\}\{-D_{q\nu\mu}^{\kappa} + (g^{\nu\mu}/\eta\omega)(\alpha_\mu^{\kappa}/\alpha_\nu^{\kappa})S_{0\kappa}^{\mu\nu}Q_{\nu\mu}^{\kappa}\}/D_q \quad (1)$$

$$\beta_{q\nu}^{\kappa} = \{(g^{\nu\nu}/\eta\omega) + (g^{\nu\mu}/\eta\omega)(\alpha_\mu^{\kappa}/\alpha_\nu^{\kappa})S_{0\kappa}^{\mu\nu}\}\{-D_{q\mu\nu}^{\kappa} + (g^{\mu\nu}/\eta\omega)(\alpha_\nu^{\kappa}/\alpha_\mu^{\kappa})S_{0\kappa}^{\nu\mu}Q_{\mu\nu}^{\kappa}\}/D_q \quad (2)$$

Here $\mu$ and $\nu$ are the band labels, $\beta_{q\mu}{}^{\kappa}$, etc. are the phonon Fourier amplitudes, $\alpha_{\mu}{}^{\kappa}$, etc. are the band fractional Fourier amplitudes, $S_{0\kappa}{}^{\mu\nu}$, $S_{\kappa}{}^{\mu\mu}$, etc. are Debye-Waller factors, $Q_{\nu\mu}{}^{\kappa}$ are mixing mode coordinates, $g^{\mu\mu}$, etc. are coupling constants, $g^{\mu\nu}$ is the mixing constant, $\eta\omega$ is the phonon quantum, $\varepsilon_{\mu}$ and $j_{\mu}$. are the local and hopping energies in $\mu$-band. The remaining definitions are:

$$D_{q\mu\nu}{}^{\kappa} = 1 + 4(j_{\mu}/\eta\omega)S_{\kappa}{}^{\mu\mu} \sin(\kappa-\Phi_{\mu\mu}{}^{\kappa}-q/2)\sin(q/2) - (g^{\mu\nu}/\eta\omega)\mathrm{Re}[(\alpha_{\nu}{}^{\kappa}/\alpha_{\mu}{}^{\kappa})Q_{\nu\mu}{}^{\kappa} S_{0\kappa}{}^{\nu\mu}] \qquad (3)$$

$$D_q = D_{q\mu\nu}{}^{\kappa} D_{q\nu\mu}{}^{\kappa} - (g^{\mu\nu}/\eta\omega)^2 |S_{0\kappa}{}^{\mu\nu}|^2 |Q_{\mu\nu}{}^{\kappa}|^2 \qquad (4)$$

$$S_{\mu\mu}{}^{\kappa} = \exp\{-(1/N)\Sigma_q|\beta_{q\mu}{}^{\kappa}|^2[1 - \cos(q)]\} \qquad (5)$$

$$\Phi_{\mu\mu}{}^{\kappa} =(1/N)\ \Sigma_q|\beta_{q\mu}{}^{\kappa}|^{2} \sin(q) \qquad (6)$$

The $\alpha$–variational equations as obtained from $\partial/\partial\alpha = 0$ are:

$$\alpha_{\mu}{}^{\kappa} = \alpha_{\nu}{}^{\kappa} \varepsilon_{\mu\nu}{}^{\kappa} / ( \Sigma_{\mu'\nu'} \alpha_{\mu'}{}^{\kappa} \alpha_{\nu'}{}^{\kappa*} \varepsilon_{\mu'\nu'}{}^{\kappa} - \varepsilon_{\mu\mu}{}^{\nu}) \qquad (7)$$

$$\alpha_{\nu}{}^{\kappa} = \alpha_{\mu}{}^{\kappa} \varepsilon_{\nu\mu}{}^{\kappa} / ( \Sigma_{\mu'\nu'} \alpha_{\mu'}{}^{\kappa} \alpha_{\nu'}{}^{\kappa*} \varepsilon_{\mu'\nu'}{}^{\kappa} - \varepsilon_{\nu\nu}{}^{\kappa}) \qquad (8)$$

$$\varepsilon_{\mu\mu}{}^{\kappa} = \varepsilon_{\mu} + 2\, j_{\mu}\, S_{\mu\mu}{}^{\kappa} \cos( \kappa- \Phi_{\mu\mu}{}^{\kappa}) + g^{\mu\mu} Q_{\mu\mu}{}^{\kappa} + \eta\omega\ (1/N)\ \Sigma_q |\ \beta_{q\mu}{}^{\kappa}\ |^2$$

$$\varepsilon_{\mu\nu}{}^{\kappa} = g^{\mu\nu} Q_{\mu\nu}{}^{\kappa} S_{0\kappa}{}^{\mu\nu} \qquad (9)$$

leading to the ground-state energy

$$E(\kappa)_{-} = ½ \{ (\varepsilon_{\mu\mu}{}^{\kappa} + \varepsilon_{\nu\nu}{}^{\kappa}) - \surd[(\varepsilon_{\mu\mu}{}^{\kappa} - \varepsilon_{\nu\nu}{}^{\kappa})^2 + 4\, \varepsilon_{\mu\mu}{}^{\kappa} \varepsilon_{\nu\nu}{}^{\kappa}]\} \qquad (10)$$

### 2.1. Starting conditions and iteration cycles

For initiating the computer calculations, the "Compaq Basic 1983" software was originally used. Later in the course of the numerical work the original results were complemented by Fortran 77 calculations and found in concert. We confined our calculations to real numbers only during the whole iteration cycle. A simple iterative program was worked out assuming appropriate starting conditions for the vibrational amplitudes, normally in the form of a "small-polaron" distribution $\beta_{q\mu}{}^{\kappa}(0)$ = const in momentum space. This phonon amplitude was used to compute zeroth-order "starting values" for the Debye-Waller factors $S_{\mu\mu}{}^{\kappa}(0)$, $S_{\nu\nu}{}^{\kappa}(0)$, $S_{\mu\nu}{}^{\kappa}$ (0), phases $\Phi_{\mu\mu}{}^{\kappa}(0)$, $\Phi_{\nu\nu}{}^{\kappa}(0)$, $\Phi_{\mu\nu}{}^{\kappa}(0$, and the mode coordinates $Q_{\mu\nu}{}^{\kappa}$. Regarded as constants these parameters were inserted into equations (1) through (6) to derive first-order vibrational amplitudes $\beta_{q\mu}{}^{\kappa}(I)$ , thereby ending up the first iterative step. The procedure was further repeated as above with $\beta_{q\mu}{}^{\kappa}(I)$ used for computing the parameters $S_{\mu\mu}{}^{\kappa}(I)$, etc. and ultimately $\beta_{q\mu}{}^{\kappa}(II)$ resulted as the former parameters were inserted into equations (1) through (6) and solved for the first-order phonon amplitudes. Such iteration steps were initiated some $N'$ = 20-30 times until the "band amplitudes" $\alpha_{\mu}{}^{\kappa}(N')$ began meeting the normalization condition

$$\sum_\mu | \alpha_\mu^\kappa |^2 = 1 \tag{11}$$

to a reasonable accuracy.

Initial checkups of our iterative method were made by applying it to the one-band Holstein Hamiltonian, viz. to the incomplete variational equations with $g^{\mu\mu} \neq 0$, $g^{\mu\nu} = g^{\nu\mu} = 0$, aimed at comparing the resulting phonon amplitudes with literature graphics, on the one hand, and at extending Merrifield's numerical studies wherever possible, on the other.

For performing the iterations as closely as possible to a "genuine" polaron distribution, Merrifield's compact forms of variational equations were used holding good at the "utility condition":

$$\cos(\kappa) > 1 - 1 / [\, 2\, (\, j_\mu / \eta\omega)\, S_{\mu\mu}^\kappa \,] \tag{12}$$

The compact forms apply to Holstein polarons only and read as follows:

$$\Delta_{\mu\mu}^\kappa = \{\, [1 + 2\, (\, j_\mu / \eta\omega\, )\, S_{\mu\mu}^\kappa \cos(\, \kappa - \Phi_{\mu\mu}^\kappa\, )]^2 - [2\, (\, j_\mu / \eta\omega)\, S_{\mu\mu}^\kappa\, ]^2 \,\}^{-3/2} \tag{13}$$

$$S_{\mu\mu}^\kappa = \exp[\, -\, (\, g^{\mu\mu} / \eta\omega\, )^2\, \Delta_{\mu\mu}^\kappa\, ] \tag{14}$$

$$\Phi_{\mu\mu}^\kappa = -\, 2\, (\, j_\mu / \eta\omega)\, (\, g^{\mu\mu} / \eta\omega)^2\, S_{\mu\mu}^\kappa \sin(\, \kappa - \Phi_{\mu\mu}^\kappa\, )\, \Delta_{\mu\mu}^\kappa \tag{15}$$

$$E(\kappa) = -\, 2\, J_\mu\, S_{\mu\mu}^\kappa \cos(\, \kappa - \Phi_{\mu\mu}^\kappa\, )\{1 - (\, g^{\mu\mu} / \eta\omega\, )^2\, \Delta_{\mu\mu}^\kappa\, \} + (\, g^{\mu\mu} / \eta\omega\, )^2\, [\, \Delta_{\mu\mu}^\kappa - 2\, (\, \Delta_{\mu\mu}^\kappa\, )^{1/3}\, ] \tag{16}$$

These forms were solved at suitable $j_\mu$ and $g^{\mu\mu}$ to generate polarons or starting momentum-space distributions $\beta_{q\mu}^\kappa$ for a single electronic band. Care was taken to keep $j_\mu \times S_{\mu\mu}^\kappa$ within the frameworks prescribed by the utility condition, since $\Delta_{\mu\mu}^\kappa = 0$ at $\cos(\kappa) < 1 - 1 / [\, 2(\, j_\mu / \eta\omega)\, S_{\mu\mu}^\kappa\, ]$.

Alternatively, Merrifield's phonon amplitudes

$$\beta_{q\mu}^\kappa = 1 / \{1 + 4(j_\mu / \eta\omega)\, S_{\mu\mu}^\kappa \sin(\, \kappa - \Phi_{\mu\mu}^\kappa - q / 2)\, \sin(q / 2)\} \tag{17}$$

were solved through iterations starting with a small-polaron distribution for calculating $S_{\mu\mu}^\kappa\,(0)$ and $\Phi_{\mu\mu}^\kappa\,(0)$. Not surprisingly, these two alternative prescriptions produced similar though not equivalent results for the optimal energies, in so far as they involved equations of very different complexity. Another problem arising from the observed inverted phases of the two respective sets of calculated quantities remains to be identified. Nevertheless, polarons or starting distributions based on the variational forms were often preferred, as they resulted from much simpler mathematical equations.

Extremal polaron energies $E(\kappa)$ obtained in this way were found in concert with published data.[4] Calculated optimal energies $E(\kappa)$, Debye-Waller factors $S_{\mu\mu}^\kappa$, Debye-Waller phases $\Phi_{\mu\mu}^\kappa$, and phonon occupation numbers $n_\mu^\kappa$ are shown in Figure 1, as obtained by solving for the phonon amplitudes and for the compact variational forms. A reasonable concord with information from published data may be seen giving credit to the iteration procedure employed presently. The finite phase seen to appear in Figure 1 is related to the asymmetric distribution of the phonon amplitudes relative to the origin in momentum space, as in the two-dimensional graphics, not shown here to avoid overfilling the file. The phonon amplitudes therein are calculated at parameters comparable to ones used earlier.[4] Throughout this paper we plot the momentum (phonon q or total $\kappa$) along the abscissa in units of $-\pi/L$ to $+\pi/L$. Most often we calculate under L = 10.

## 2.2. Vibronic polarons

Before all, we consider it reasonable to introduce a few definitions. Having in mind the traditional regime of parameters used for polaron research, we divide the electron energy axis into three ranges as follows:

- Low-energy range where $|e_{\mu\nu}| / \eta\omega \ll 1$
- Intermediate-energy range where $|e_{\mu\nu}| / \eta\omega \sim 1$
- High-energy range where $|e_{\mu\nu}| / \eta\omega \gg 1$,

where $|e_{\mu\nu}|$ is the PJT gap energy or any other energy-dependent electronic parameter, such as the coupling constants and the hopping energies. The reasoning behind the above definitions comes from the adiabatic effect being expandable in the powers of $\eta\omega / |e_{\mu\nu}|$ which enters as a "small parameter". We designate them as ranges I through III in the order of increasing electron energy. Range III is traditional for polaron studies by means of the adiabatic approximation. Unlike it, ranges I and II have not been explored so far even though doubts have been raised as to the applicability of adiabatic methods to these ranges [9]. Solving for the (nearly) complete variational equations (1) through (6) with (nearly) all the three electron-phonon coupling constants finite (we remind that because of band-interchange symmetry one usually assumes $g^{\mu\nu} = g^{\nu\mu}$ reducing the number of coupling constants by one) proved less easy, apparently due to a poorer convergence of the iterational cycles relative to the one-band solutions. The main reason for that was apparently in the finite $g^{\mu\nu}$ term which resulted in a higher complexity of the variational solutions and the related higher risk of the program getting overloaded during a cycle. These complications resulted in the necessity of operating at small numerical values of the parameters $j_\mu$, $g^{\mu\mu}$, $g^{\mu\nu}$, etc. in fractions of a phonon quantum $\eta\omega$. This program limitation brought our polaron calculations to the strongly nonadiabatic regime of range I, $j_\mu$ , $|e_{\mu\nu}| \ll \eta\omega$ ,which has not been thoroughly explored so far. An additional though always standing limitation was the requirement to operate in real numbers during the whole iteration cycle. We should also remind that seeking extremal solutions in real numbers does not preclude performing some of the iterations on the complex plane, which is not the case presently.

Nevertheless, with the foregoing range I in mind, we define the following polaron types, depending on the relative values of the remaining parameters at $G_{\mu\nu} \neq 0$ (cf. [2],[10]):

- Adiabatic polaron (AD) for $G_{\mu\nu}^2 < 2 J_\mu$ ,
- Antiadiabatic polaron (AAD) for $G_{\mu\nu}^2 > 2 J_\mu$ ,
- Jahn-Teller polaron (JT) for $E_{\mu\nu} = 0$ ,
- Pseudo-Jahn-Teller polaron (PJT) for $E_{\mu\nu} \neq 0$ ,
- Weakly coupled polaron (WC) for $2G_{\mu\nu}^2 < E_{\mu\nu}$ ,

- Strongly coupled polaron (SC) for $2G_{\mu\nu}^2 > E_{\mu\nu}$ ,
- Di-coupled polaron (DC) for $G_{\mu\mu} = 0$, $G_{\nu\nu} \neq 0$ ,
- Tri-coupled polaron (TC) for $G_{\mu\mu} \neq 0$, $G_{\nu\nu} \neq 0$ ,
- Semibound polaron (SB) for $J_\mu = 0$, $J_\nu \neq 0$ ,
- Bound polaron (B) for $J_\mu = 0$, $J_\nu = 0$ .

We made a number of specific calculations on various polaron types in the nonadiabatic regime which will be presented next in Figures. Most of the vibronic-polaron calculations were carried out under starting conditions close to real distributions for the phonon amplitudes. For that purpose, the one-band Merrifield equations were used for generating starting conditions in the form of a Holstein polaron for either component band of the vibronic problem. At the same time, the band amplitudes started with a pre-selected value $\alpha_\mu^\kappa (0) = 1 / \sqrt{2}$, the equivalent band contribution. Checks were made to ensure that the resulting phonon amplitudes were not affected by any deliberate change of the starting conditions beyond an acceptable amount of only a few percent. All calculations were made at the phonon energy of 0.01 eV.

Under these conditions, the vibronic variational equations were found to generate *small polarons* with phonon amplitudes independent of the wave number q in phonon momentum space, within the parameter range I overlapping some of range II, namely $G_{\mu\nu} \leq 0.5$, where the program operated properly. This numerical conclusion may not be surprising because of the expected size of the configurational distortion, due to the strongly local character of the band-mixing by phonons, especially in the lower-energy electronic states.

Here and elsewhere, we express all relevant parameters in units of a phonon quantum, viz. $G_{\mu\nu} = g^{\mu\mu} / \eta\omega$, $J_\mu = j_\mu / \eta\omega$, $E_{\mu\nu} = |e_{\mu\nu}| / \eta\omega$, etc. Figure 2 shows the E(κ) diagrams for AD JT and AD PJT polarons. The phonon dressing effect for these polarons being slight, it possibly brings the convergency of the iteration cycles to a low, as suggested by the higher roughness of the diagrams. Indeed, the E(κ) diagrams smoothen on passing to AAD JT and AAD PJT polarons, as seen below in Figure 4. In both Figures 2 and 3, the dressing effect is also seen to increase, in as much as the polaron bandwith drops, from JT to PJT. This is an essential conclusion, in so far as it signifies that the Pseudo-Jahn-Teller polaron involves more phonons than does the Jahn-Teller polaron. Consequently, PJT polarons could be expected to be heavier and less mobile. Calculating the vibronic polaron mass will be dealt with elsewhere.

In Figure 4 (a) through (j), we depict the remaining quantities for AAD PJT polarons, so as to demonstrate the program capacity, namely the DW factors $S_{\mu\mu}^\kappa$, $S_{\nu\nu}^\kappa$, $S_{\mu\nu}^\kappa$, mixing-mode coordinate $Q_{\mu\nu}^\kappa$, phonon-occupation numbers $n_\mu^\kappa$, $n_\nu^\kappa$, band amplitudes $\alpha_\mu^\kappa$, $\alpha_\nu^\kappa$, and the phonon amplitudes $\beta_{q\mu}^\kappa$, $\beta_{q\nu}^\kappa$. We remind that the phonon amplitudes of all vibronic polarons studied presently were found of the small-polaron character, that is, independent of the phonon momentum q and symmetric relative to inversion along the q-axis. Under these conditions, the DW phases are both vanishing: $\Phi_{\mu\mu}^\kappa \equiv 0$, $\Phi_{\nu\nu}^\kappa \equiv 0$. We see that there are more phonons along the κ-axis towards the Brillouin-zone center in μ-band, as evidenced by $S_{\mu\mu}^\kappa$ and $n_\mu^\kappa$, and towards the Brillouin-zone edges in ν-band, as evidenced by $S_{\nu\nu}^\kappa$ and $n_\nu^\kappa$ . The absolute mixing- mode coordinate $Q_{\mu\nu}^\kappa$ is more significant towards the zone center, unlike the mixing DW factor which is more significant towards the zone edges. The μ-band contributes to a great deal towards the zone edges, while the ν-band does so towards the zone center. Also, the distribution of phonon amplitudes is complementary in $\mu$– and $\nu$–band. They are both characteristic of a not flat phonon distribution in $\kappa$–space.

Most the above examples have described features in $\kappa$–space of small polarons itinerant in both electronic bands. From the viewpoint of scientific interest, however, it would be important to see what happens if carriers in one band are immobilized, e.g. due to a vanishing bandwidth in electron ground state. At the same time, the bandwidth in electron excited state may be finite. The resulting polaron moves by virtue of the ground-state to excited-state mixing by phonons. In the adiabatic case, the poorer convergency was again manifested unambiguously, though not shown. Examples of $E(\kappa)$ diagrams of Jahn-Teller and Pseudo-Jahn Teller semibound antiadaiabatic polarons (SB AAD) are shown next in Figure 5. The band-narrowing from JT to PJT is particularly clear in this case. In Figure 6 (a) through (e), we display the $\kappa$–space distributions of the quantities relevant to the Jahn-Teller SB AAD polaron of Figure 5. In (a), we show the diagonal DW factors, $S_{\mu\mu}{}^{\kappa}$ and $S_{\nu\nu}{}^{\kappa}$. The off-diagonal DW factor $S_{\mu\nu}{}^{\kappa}$ exhibits an unique double-wave character. This double-wave character does not appear incidental, since it is also exhibited by the mixing-mode coordinate $Q_{\mu\nu}{}^{\kappa}$ in (b). The momentum distribution of the phonon occupation numbers $n_{\mu}{}^{\kappa}$ and $n_{\nu}{}^{\kappa}$ as well as of the band amplitudes $\alpha_{\mu}{}^{\kappa}$, $\alpha_{\nu}{}^{\kappa}$ are shown in (c) and (d), respectively. Finally in (e), we plot the momentum distributions of the phonon amplitudes $\beta_{q\mu}{}^{\kappa}$, $\beta_{q\nu}{}^{\kappa}$.

We also studied multi-coupled vibronic polarons to check if these could survive within the present variational frames. Results are shown in Figure 7 where we plot the $E(\kappa)$ diagram for a tri-coupled Jahn-Teller polaron at parameters immitating a previous calculation [8]. By virtue of the finite interband mixing constant $G_{\mu\nu}$, this species does not collapse into a single-band Holstein polaron.

Finally, we dealt with fully bound Pseudo-Jahn-Teller polarons with both band-hopping energies $J_{\mu}$ and $J_{\nu}$ vanishing. Under localized conditions, there are no $\kappa$–space distributions which now reduce to a single value. Table I lists calculated quantities relevant to: strong coupling ($2G_{\mu\nu}{}^2 > E_{\mu\nu}$), critical coupling ($2G_{\mu\nu}{}^2 = E_{\mu\nu}$), and weak coupling ($2G_{\mu\nu}{}^2 < E_{\mu\nu}$), all the three criteria being borrowed from the analytic adiabatic approximation. We stress the behavior of the absolute mixing-mode coordinate which is seen to increase from strong to weak coupling, in contrast with the adiabatic result. An apparent explanation should perhaps be sought in the present extreme range of parameters where the adiabatic approximation simply does not hold good.

## 3. Conclusion

Hopefully, this investigation is intended to be the first of a series dealing with the variational band theory of vibronic polarons. Because of limitations imposed by both calculational language and program, our study was confined to the real-numbers axis. This reduced our capacity considerably, in so far as some of the iteration cycle may normally go onto the complex plane, even if the final result is obtained in real numbers. The form of variational equations derived for vibronic polarons suggests that excluding the complex plane may be the primary reason behind the program being unable to cover the energy range beyond the present range I (fraction of a phonon quantum) regime. Nevertheless, the latter regime is worth studying in itself, since it overlaps with an essential portion of the "nonadiabaticity range" for which little if anything is known, since the adiabatic approximation, the traditional source of analytic conclusions, does not hold true therein. For this reason, we believe the present investigation does have a heuristic value in raising several points of scientific importance, to be discussed below.

The phonon dressing effect is demonstrated to be light in the low-energy range I. Our vibronic variational equations are found to generate small polarons, strongly confined in real space. This numerical conclusion appears correct in view of the expected size of configurational distortion, arising from the strongly local character of the band mixing by phonons. Another intriguing observation is that the polaron effect increased as it should from adiabatic to antiadiabatic and from Jahn-Teller to Pseudo-Jahn-Teller. We also consider it non-trivial that stable semi-bound vibronic polarons are found to generate, that can have a profound effect on our understanding of the behavior of carriers trapped in ground electronic state which may, however, migrate if in the excited electronic state: the ground state-excited state mixing effect makes them all itinerant. Stable di-coupled and tri-coupled polarons are prohibited by group theory on grounds of the incompatible mode symmetries in that $g^{\mu\nu}$ and $g^{\mu\mu}$ may not be both finite for a vibrational mode of a given symmetry. Nevertheless, we find these polarons firmly itinerant raising the hope that certain selection-breaking compromise may eventually be found in experimentally important cases. On the other hand, our variational conclusions for bound polarons may not be compared directly against the strong background of analytic data, since the analytic adiabatic approximation simply may not apply at gap energies below a phonon quantum, which is the present vibronic parameter range I.

Further work is planned to incorporate the complex numbers so as to extend the regime to larger values of the vibronic parameters, possibly to the adiabatic energy range III. This would make it possible to compare variational results with analytic results obtained by using the adiabatic approximation. An important adiabatic result missing presently is the transition from small to large bound PJT polarons, as the energy gap $E_{\mu\nu}$ increases towards the critical value of $2G_{\mu\nu}^2$ at constant mixing constant. Indeed, this text-book conclusion has been drawn for the adiabatic energy range III. Another one is the not-adiabatic behavior of the mixing-mode coordinate $Q_{\mu\nu}$ of bound JT polarons which is found to increase as the gap increases, in contrast to the range III result. On this ground, it is perhaps not surprising why our solutions for the non-adiabatic range I yield small polarons only. We see that linking the energy ranges I and III is of chief importance for understanding the vibronic polarons.

## 4. References

[1] M. Georgiev, cond-mat/0601676.
[2] D.W. Brown, K. Lindenberg and M. Georgiev, cond-mat/0602052.
[3] R.E. Merrifield, J. Chem. Phys. **40**, 445-450 (1964).
[4] Y. Zhao, D.W. Brown and K. Lindenberg, J. Chem. Phys. **106**, 5622-5630 (1997).
[5] See A.G. Andreev, S.G. Tsintsarska, M.D. Ivanovich, I. Polyanski, M. Georgiev and A.D. Gochev, Central Eur. J. Phys. **2**, 329-356 (2004).
[6] A.J. Milne, Nature **392**, 147 (1998).
[7] I.B. Bersuker and V.Z. Polinger,: *Vibronic interactions in molecules and crystals* (Springer, Berlin, 1989).
[8] K.-H. Höck, H. Nickisch and H. Thomas, Helvetica Phys. Acta **56**, 237-243 ( 1983).
[9] M. Wagner, J. Chem. Phys. **82**, 3207 (1984).
[10] L.J. De Jongh, Physica C **152**, 171-216 (1988).

Table I

Calculated data for bound vibronic polarons

(a) Strongly coupled bound polaron
(b) Critically coupled bound polaron
(c) Weakly coupled bound polaron

| Quantity | (a) | (b) | (c) |
|---|---|---|---|
| $J_\mu$ | 0 | 0 | 0 |
| $J_\nu$ | 0 | 0 | 0 |
| $G^{\mu\mu}$ | 0 | 0 | 0 |
| $G^{\nu\nu}$ | 0 | 0 | 0 |
| $G^{\mu\nu}$ | 0.1 | 0.1 | 0.1 |
| $E_{\mu\nu}$ | 0.01 | 0.02 | 0.04 |
| $E_{min}$ | -0.000128 | -0.000171 | -0.000182 |
| $S_{\mu\mu}$ | 0.999205 | 0.999742 | 0.999985 |
| $S_{\nu\nu}$ | 0.878104 | 0.704710 | 0 |
| $S_{\mu\nu}$ | 0.951037 | 0.860226 | 0 |
| $Q_{\mu\nu}$ | -0.379812 | -0.593664 | -5.536731 |
| $n_{\mu\mu}$ $n_{\nu\nu}$ | 0.000760 0.124081 | 0.000246 0.334061 | 0.000014 30.61343 |
| $\alpha_\mu$ $\alpha_\nu$ | 0.908629 0.442809 | 0.947858 0.306350 | 0.935491 0.553014 |
| $\alpha_\mu^2 + \alpha_\nu^2$ | 1.021687 | 0.992285 | 1.180969 |
| $\beta_{q\mu}$ $\beta_{q\nu}$ | -0.02756 -0.35225 | -0.01568 -0.57798 | -0.00379 -5.53294 |

## 5. Figure captions

Figure 1. Comparing numerical calculations for the Merrifield-Holstein polaron by the standard variational equations (1) through (10) (s) with ones by the compact variational forms (13) through (16) (c), as follows:
(a) Debye-Waller amplitudes $S_{\mu\mu}(\kappa)$,
(b) Debye-Waller phase $\Phi_{\mu\mu}(\kappa)$,
(c) extremal polaron energies $\Delta E(\kappa) = E(\kappa) - E(0)$, and
(d) phonon occupation numbers $n_\mu(\kappa)$ and compact variational forms $\Delta_{\mu\mu}(\kappa)$.
All these are plotted versus the total crystalline momentum $\kappa$. Note the inverted phases of respective resulting quantities by the two methods. The following parameters are used: $J_\mu = 0.1$ and $G_{\mu\mu} = -1$ (in units of a phonon quantum), $\varepsilon_{\mu\mu} = 0$, and $\eta\omega = 0.01$ eV.

Figure 2. Calculated $E(\kappa)$ diagrams for Jahn-Teller and Pseudo-Jahn-Teller adiabatic polarons. The polaron bandwidth apparently tends to narrow as one goes from JT to PJT. The parameters are: $J_\mu = 0.03$, $J_\nu = 0.06$, $G_{\mu\nu} = 0.1$, $E_{\mu\nu} = 0.01$, $\eta\omega = 0.01$ eV.

Figure 3. Same as Figure 2 for Jahn-Teller and Pseudo-Jahn-Teller antiadiabatic polarons. The JT to PJT band narrowing is now even more pronounced. The parameters are: $J_\mu = 0.00003$, $J_\nu = 0.00006$, $G_{\mu\nu} = 0.1$, $E_{\mu\nu} = 0.01$, $\eta\omega = 0.01$ eV.

Figure 4. Depicting the remaining quantities for Pseudo-Jahn-Teller antiadiabatic polarons from the data leading to the latter case in Figure 3, as follows:
(a) $S_{\mu\mu}{}^{\kappa}$, (b) $S_{\nu\nu}{}^{\kappa}$, (c) $S_{\mu\nu}{}^{\kappa}$, (d) $Q_{\mu\nu}{}^{\kappa}$, (e) $n_\mu{}^{\kappa}$, (f) $n_\nu{}^{\kappa}$, (g) $\alpha_\mu{}^{\kappa}$, (h) $\alpha_\nu{}^{\kappa}$, (i) $\beta_\mu{}^{\kappa}$, (j) $\beta_\nu{}^{\kappa}$.

Figure 5. Calculated $E(\kappa)$ diagram for Jahn-Teller and Pseudo-Jahn-Teller semibound antiadiabatic polarons. The band-narrowing from JT to PJT is clearly seen. The parameters are: $J_\mu = 0$, $J_\nu = 0.002$, $G_{\mu\nu} = 0.1$, $E_{\mu\nu} = 0.01$, $\eta\omega = 0.01$ eV.

Figure 6. Calculated momentum-space distributions of the quantities relevant to the Jahn-Teller semibound antiadiabatic polaron of the preceding Figure. We plot:
(a) the DW factors, diagonal $S_{\mu\mu}{}^{\kappa}$ and $S_{\nu\nu}{}^{\kappa}$ and off-diagonal $S_{\mu\nu}{}^{\kappa}$,
(b) the mixing-mode coordinate $Q_{\mu\nu}{}^{\kappa}$,
(c) the phonon occupation numbers $n_\mu{}^{\kappa}$ and $n_\nu{}^{\kappa}$,
(d) the band amplitudes $\alpha_\mu{}^{\kappa}$ and $\alpha_\nu{}^{\kappa}$ and,
(e) the phonon amplitudes $\beta_\mu{}^{\kappa}$ and $\beta_\nu{}^{\kappa}$.
The double-wave character of the off-diagonal DW factor $S_{\mu\nu}{}^{\kappa}$ and the phonon coordinate $Q_{\mu\nu}{}^{\kappa}$ are to be distinguished.

Figure 7. Calculated $E(\kappa)$ diagram for a Jahn-Teller tri-coupled polaron in the adiabatic regime. The parameters are: $J_\mu = 0.02$, $J_\nu = 0.04$, $G_{\mu\mu} = -0.1$, $G_{\nu\nu} = 0.1$, $G_{\mu\nu} = 0.1$, $E_{\mu\nu} = 0$, $\eta\omega = 0.01$ eV. By virtue of the finite interband mixing constant $G_{\mu\nu}$, this species does not collapse into a single-band Holstein polaron.

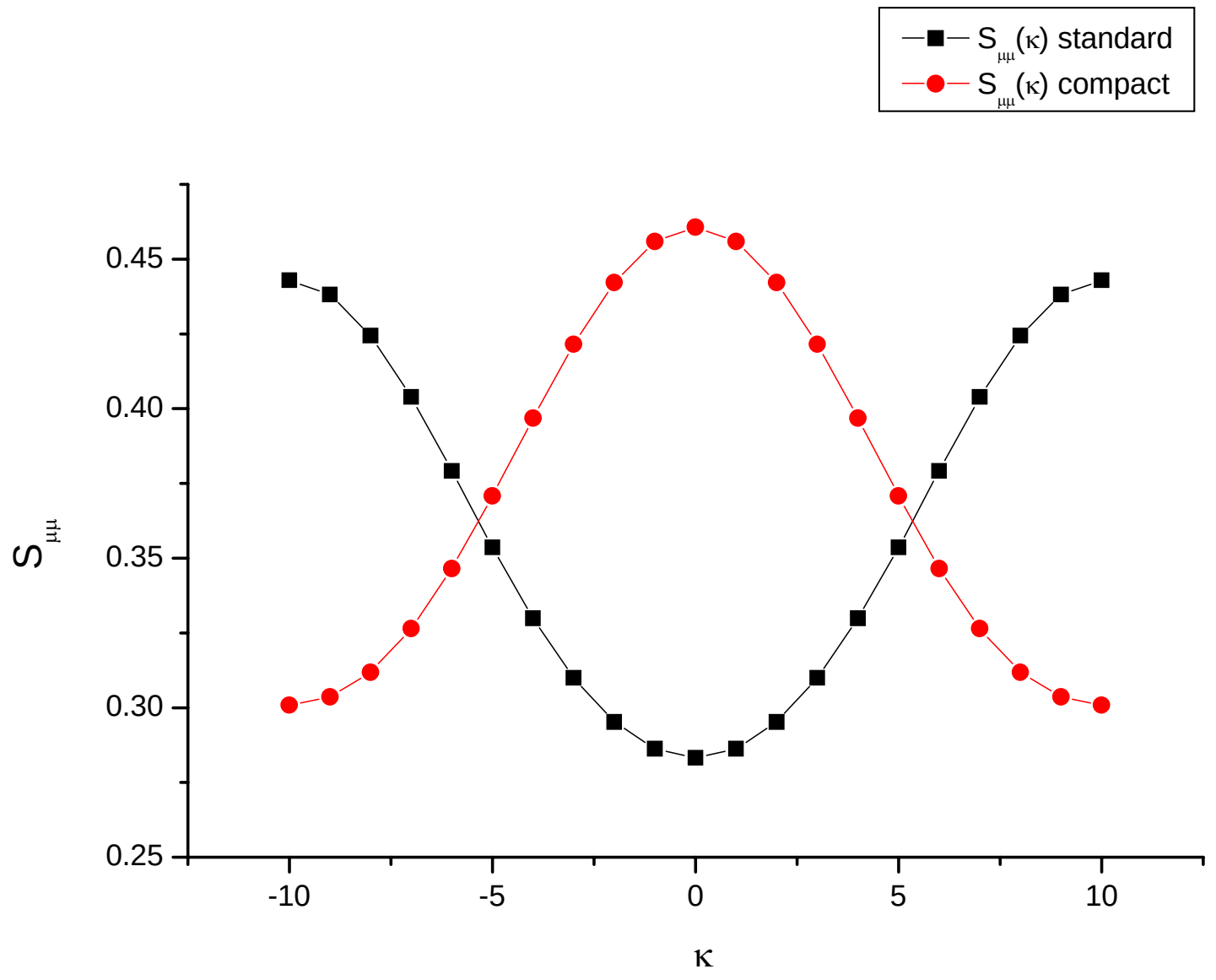

Figure 1 (a)

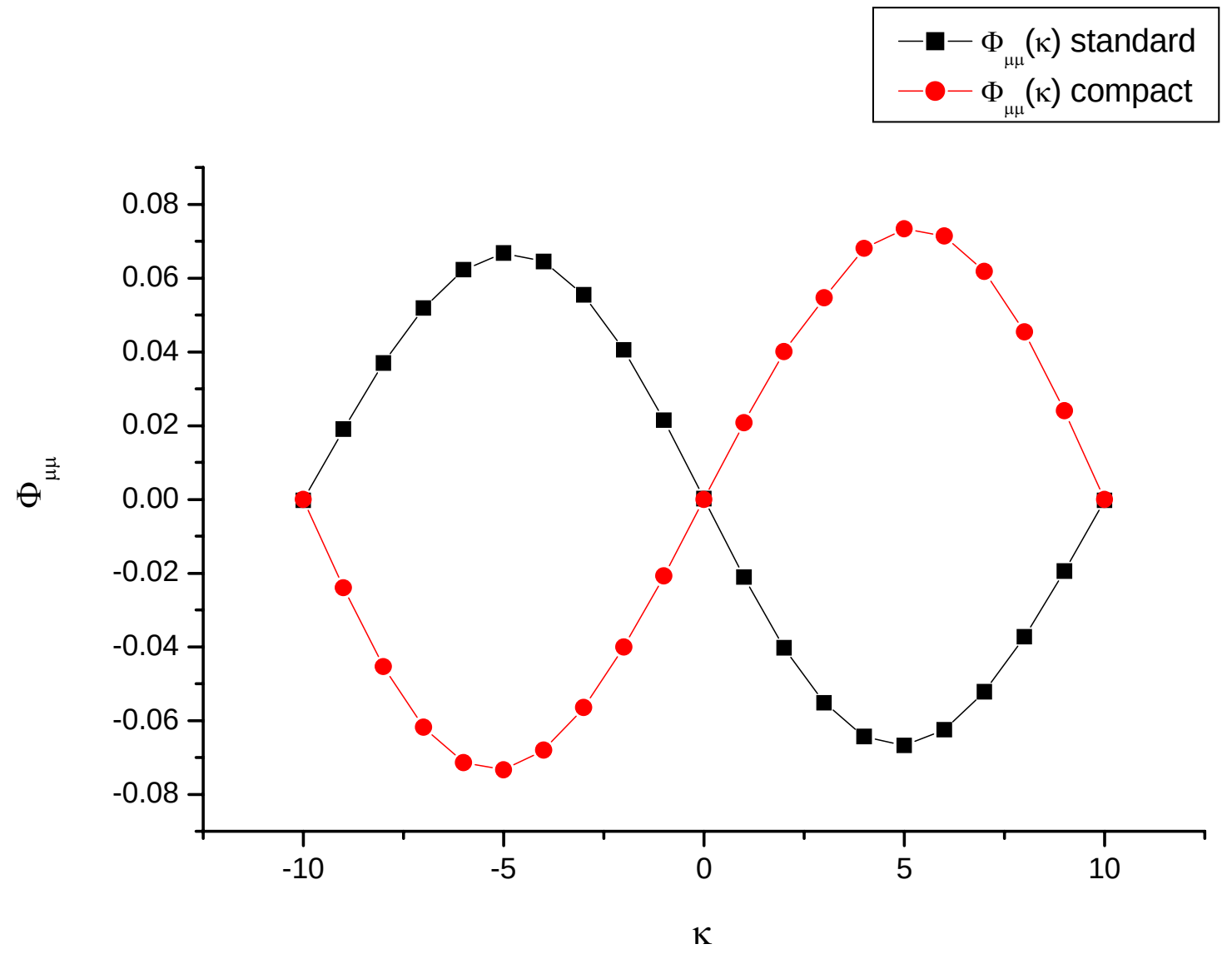

Figure 1 (b)

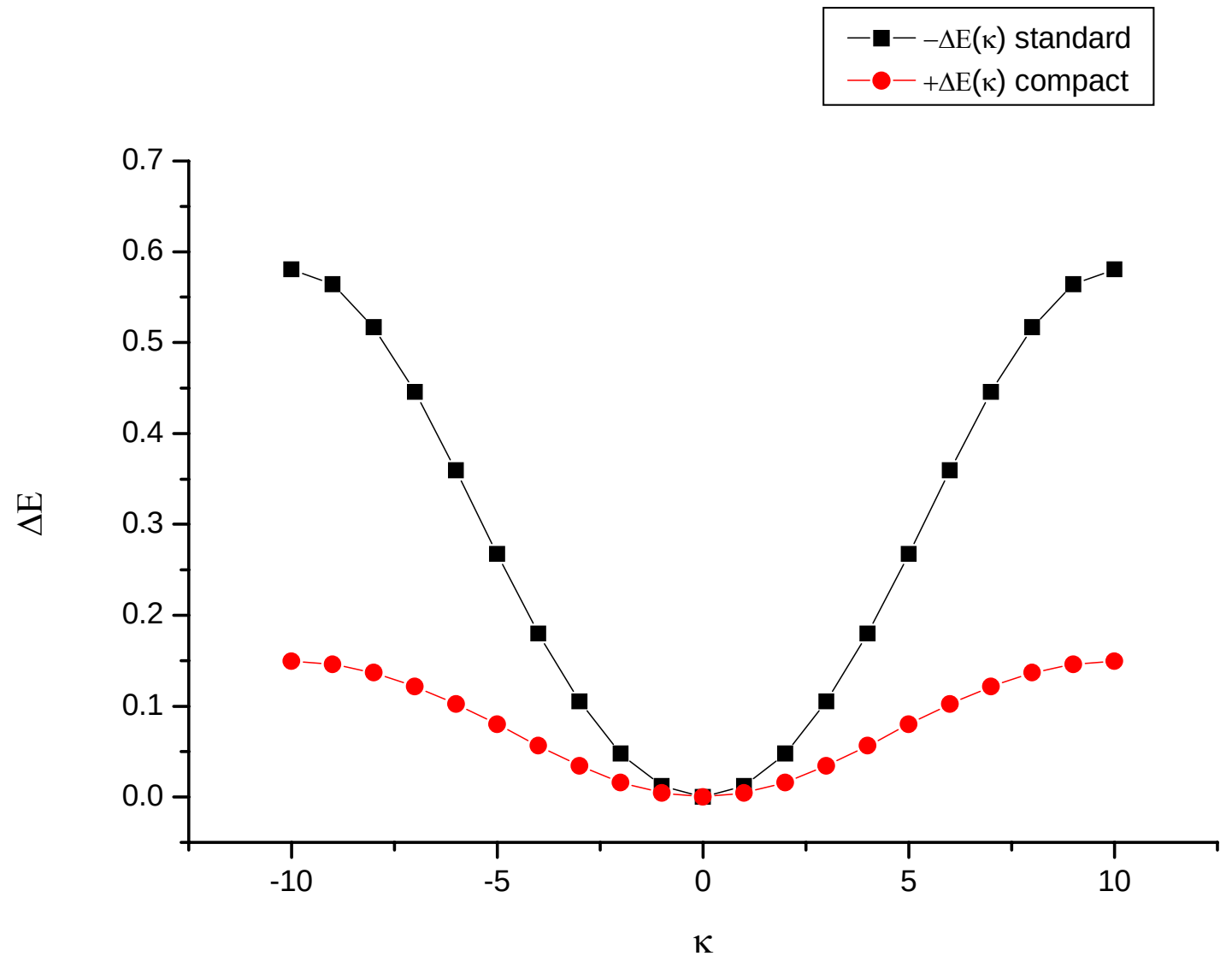

Figure 1 (c)

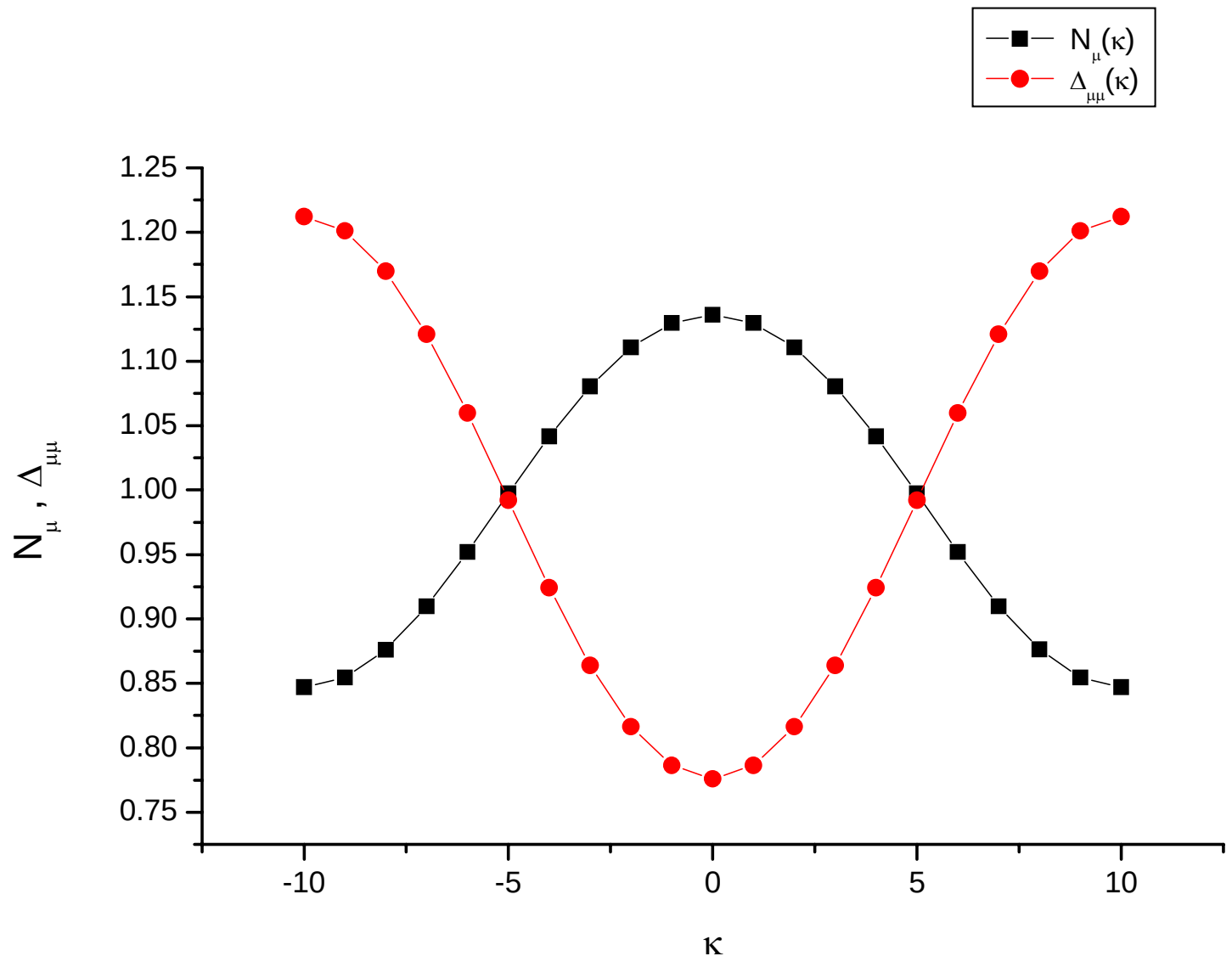

Figure 1 (d)

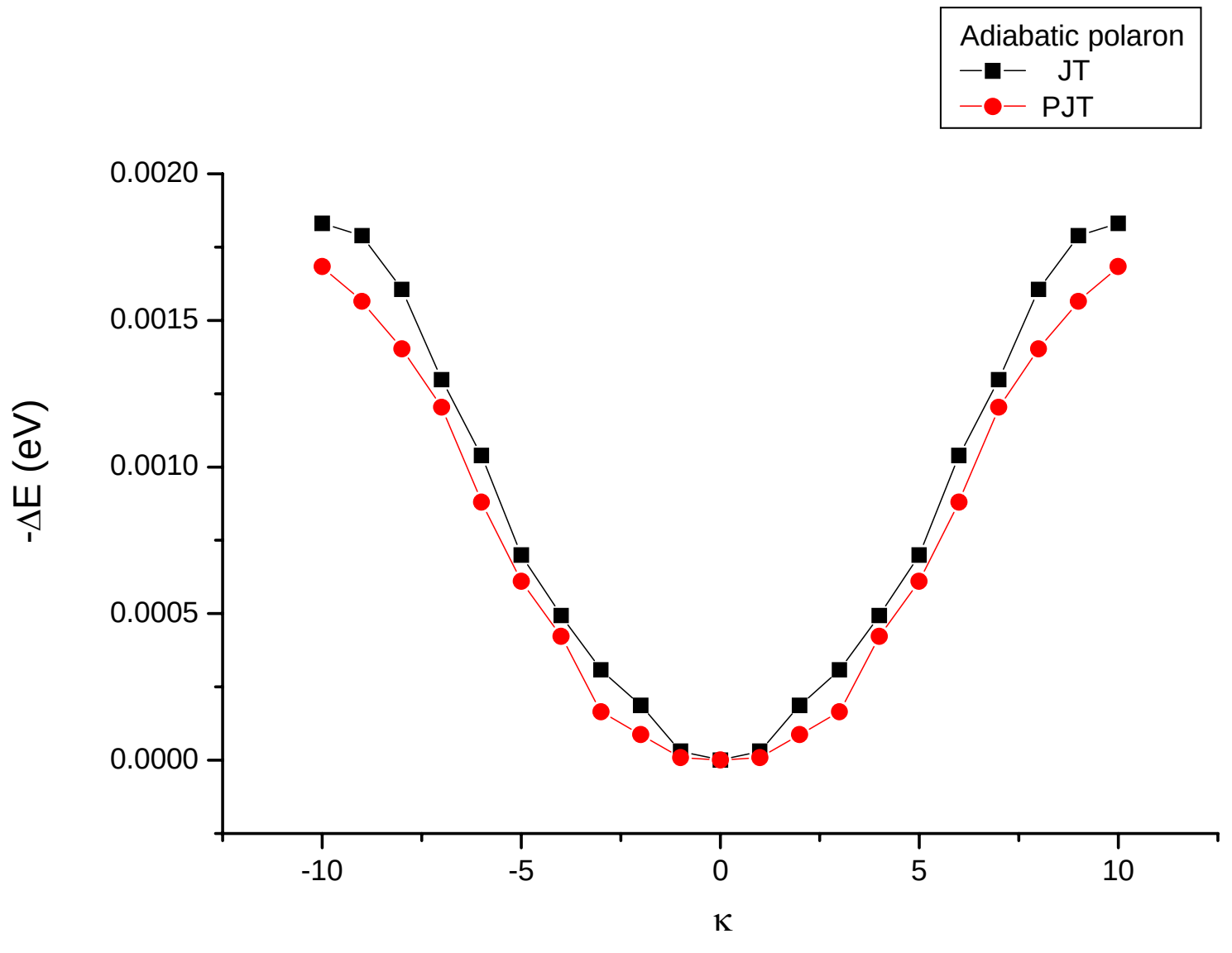

Figure 2

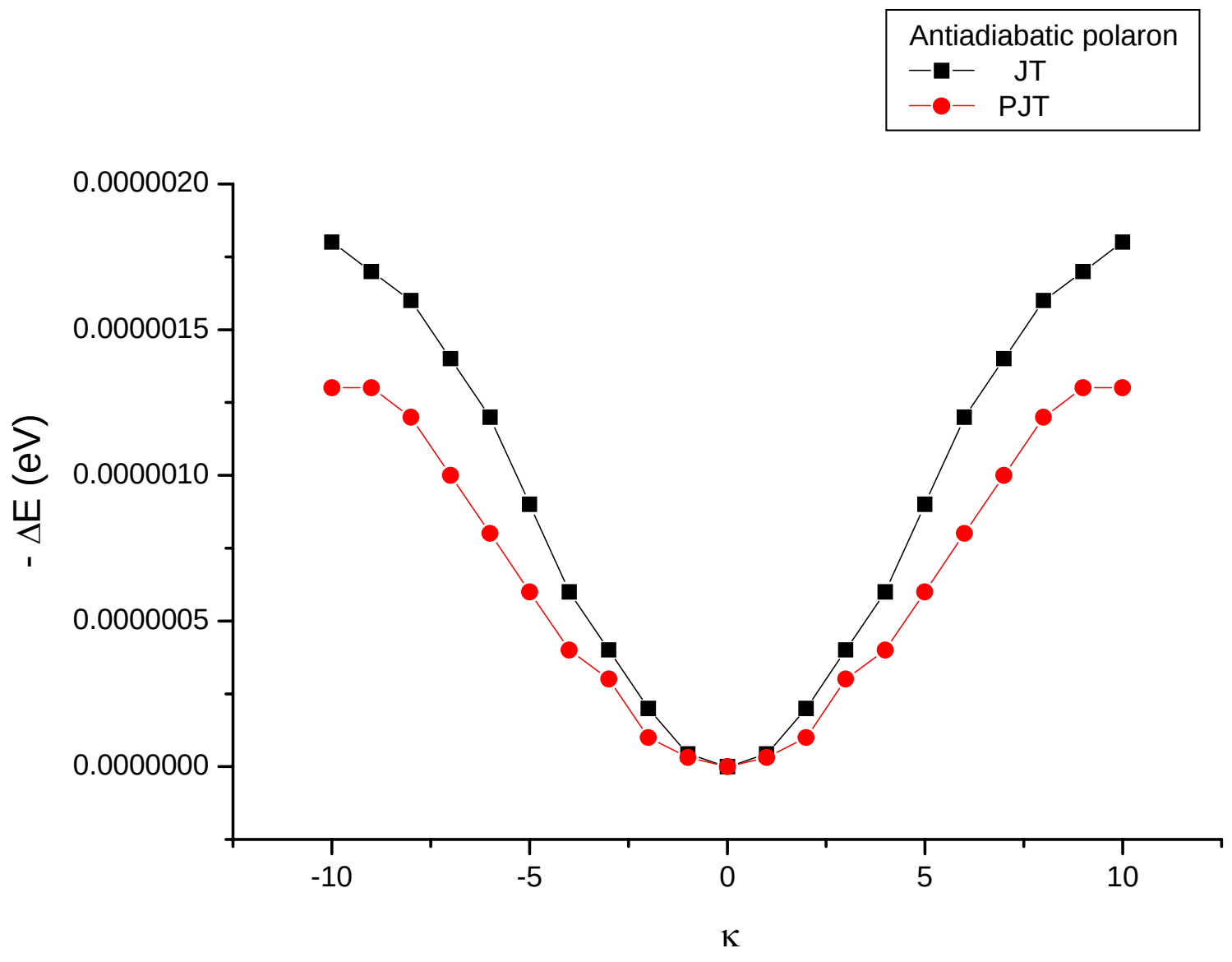

Figure 3

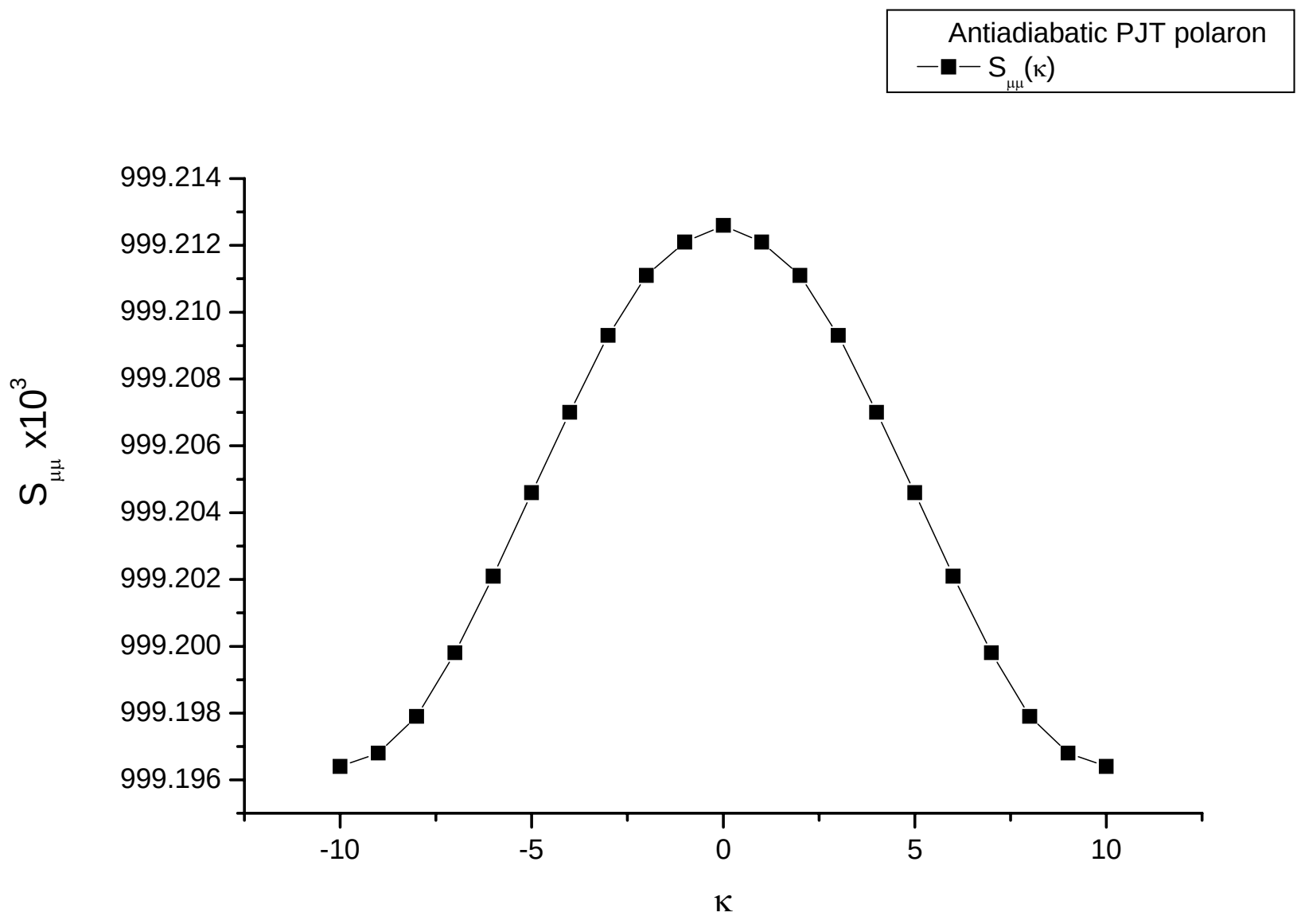

Figure 4 (a)

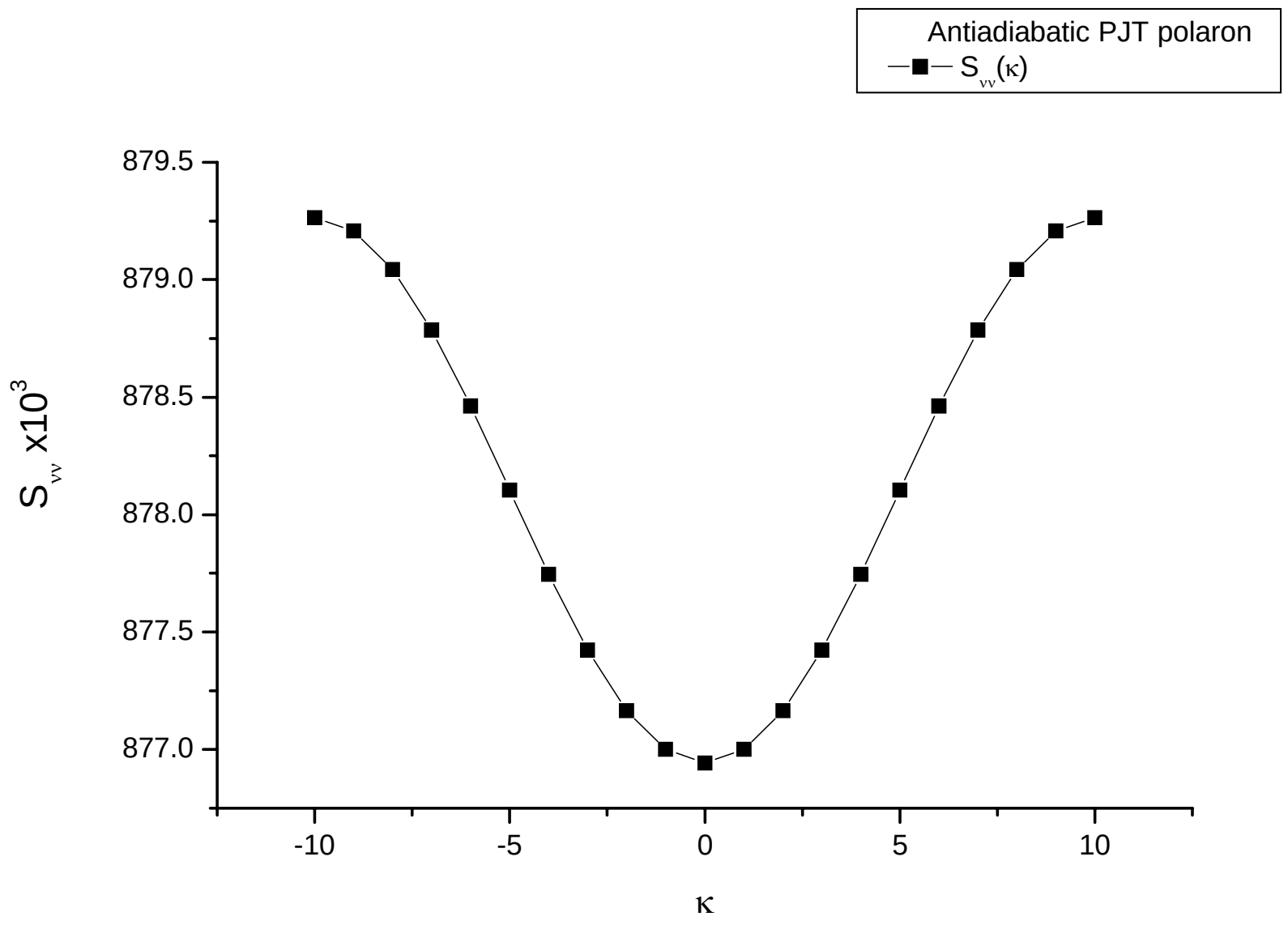

Figure 4 (b)

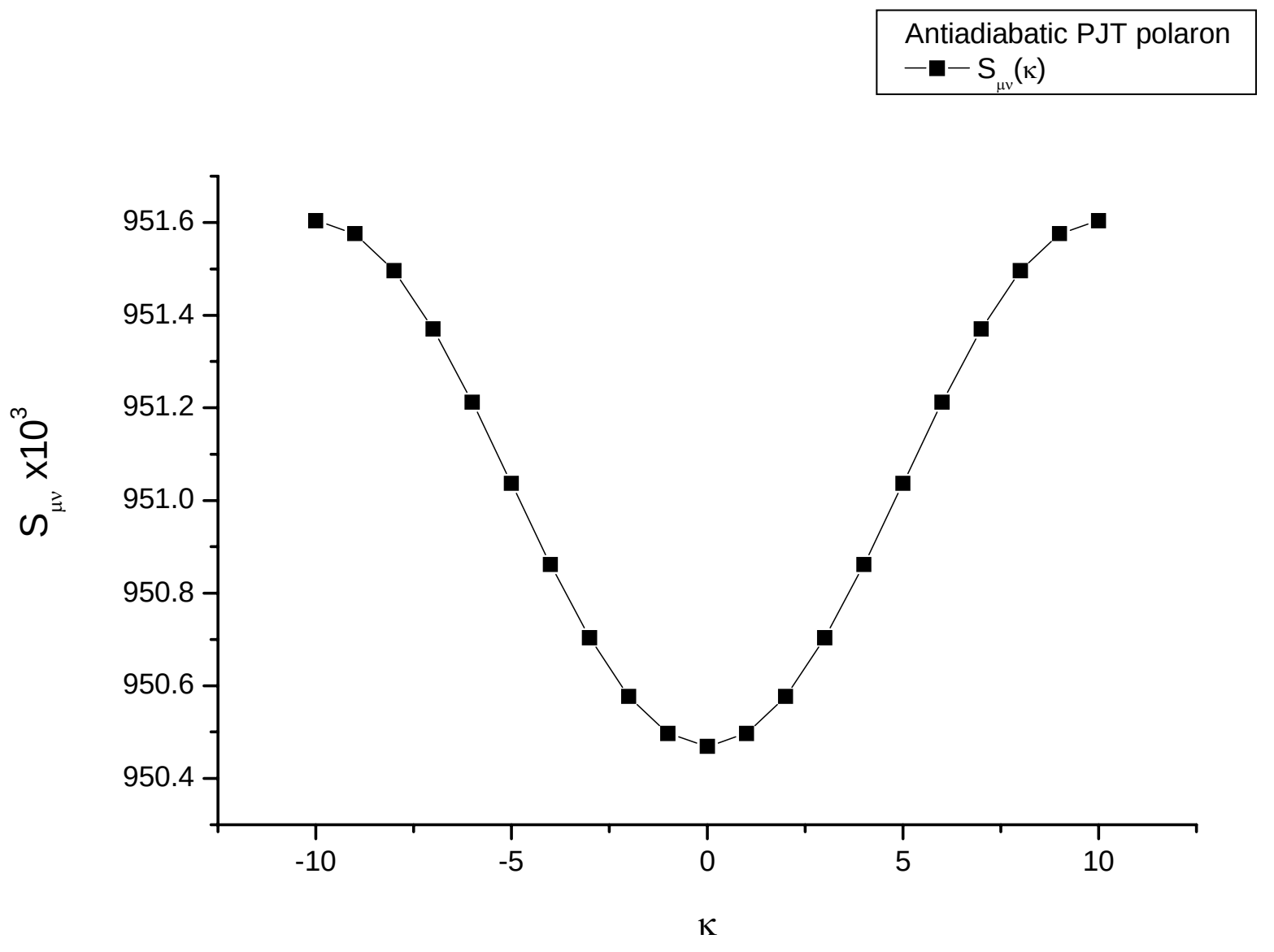

Figure 4 ( c)

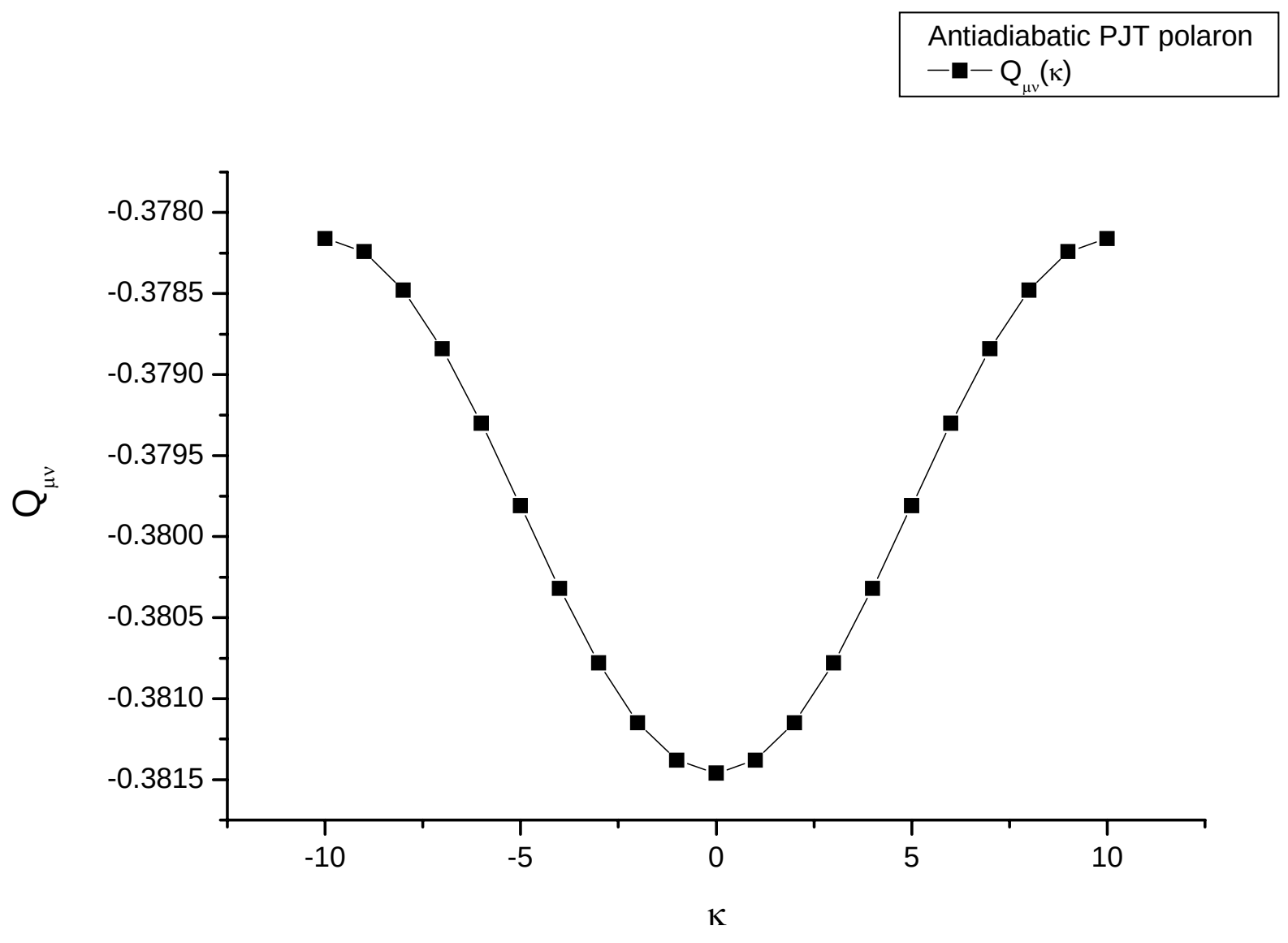

Figure 4 (d)

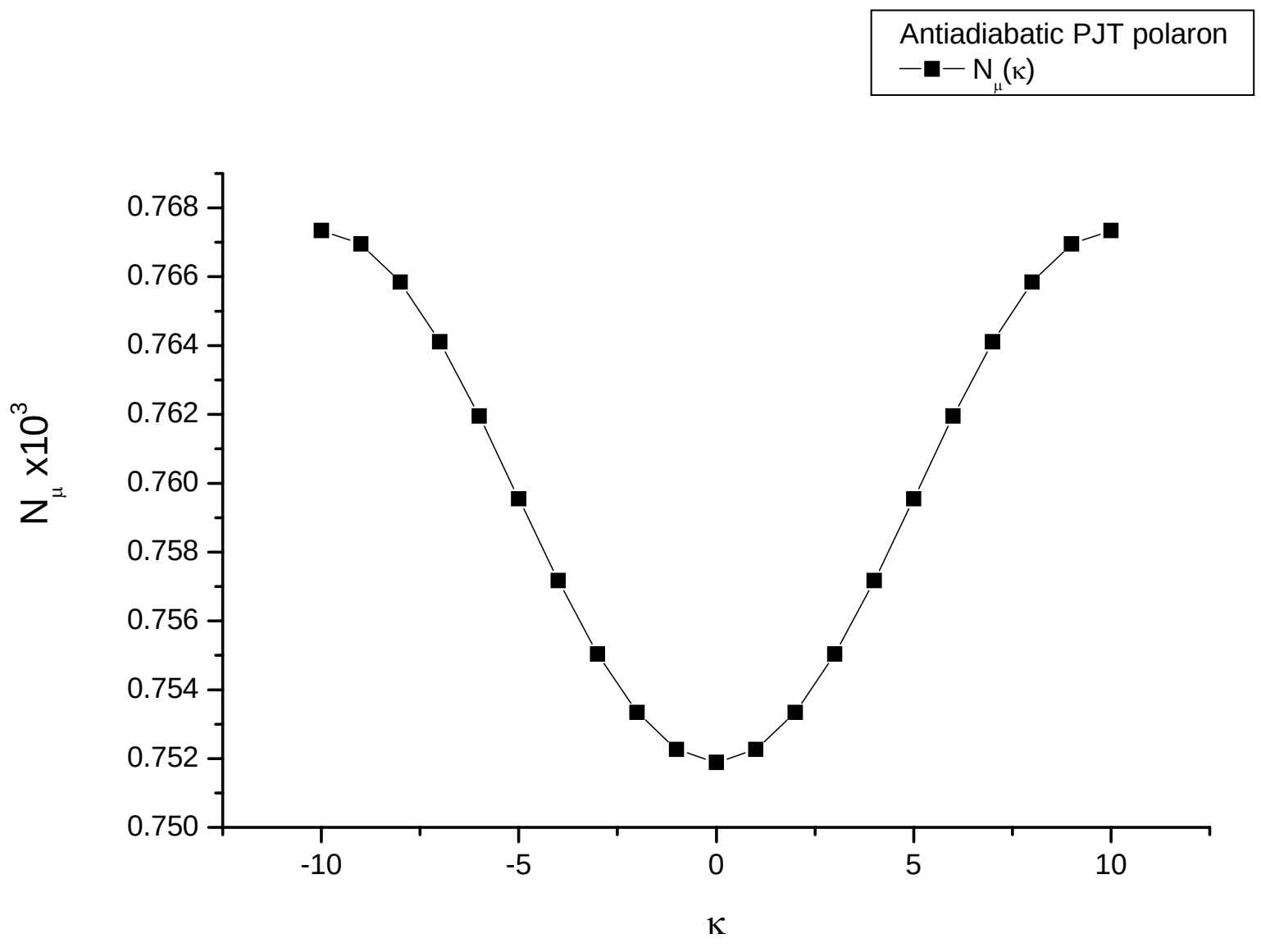

Figure 4 (e)

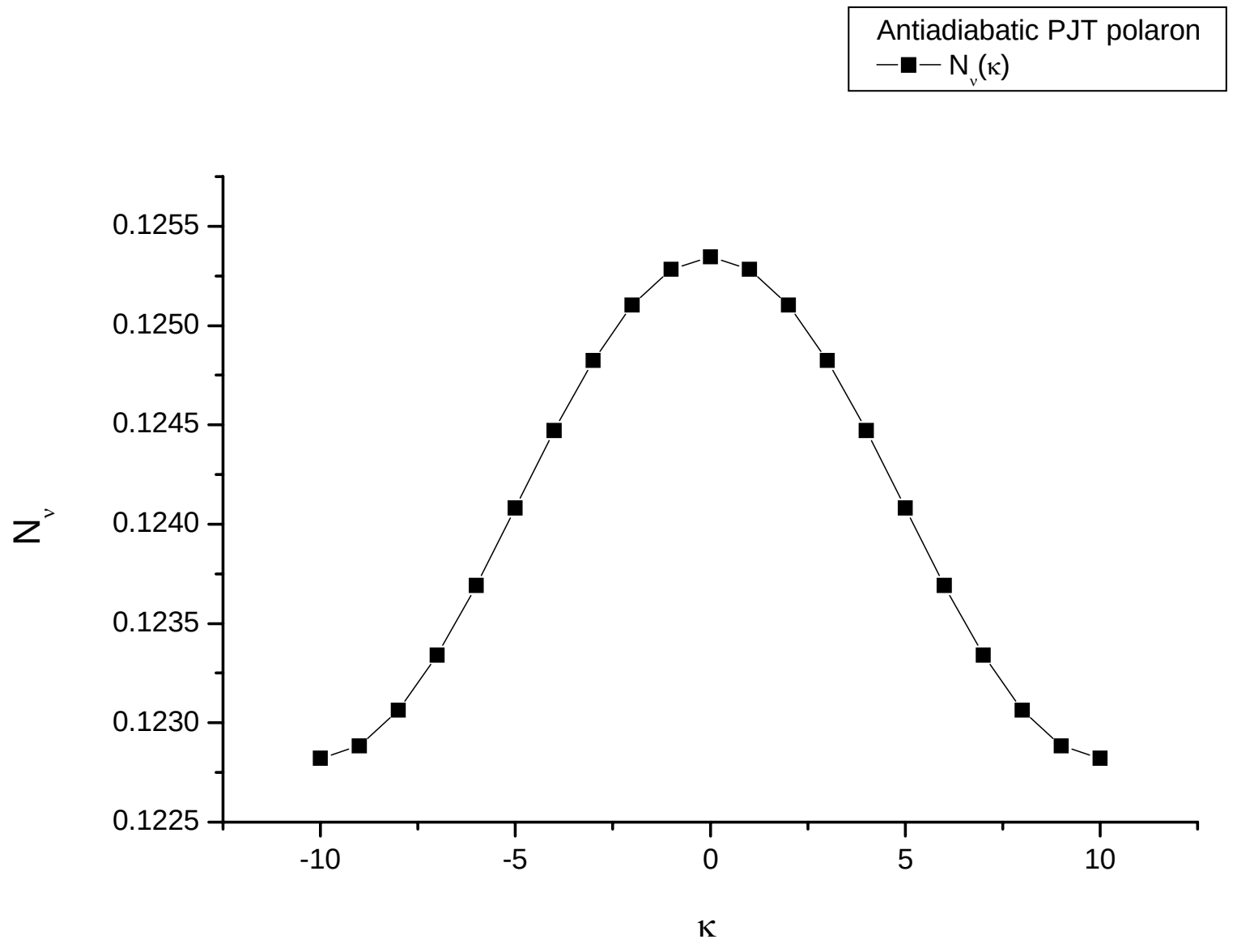

Figure 4 (f)

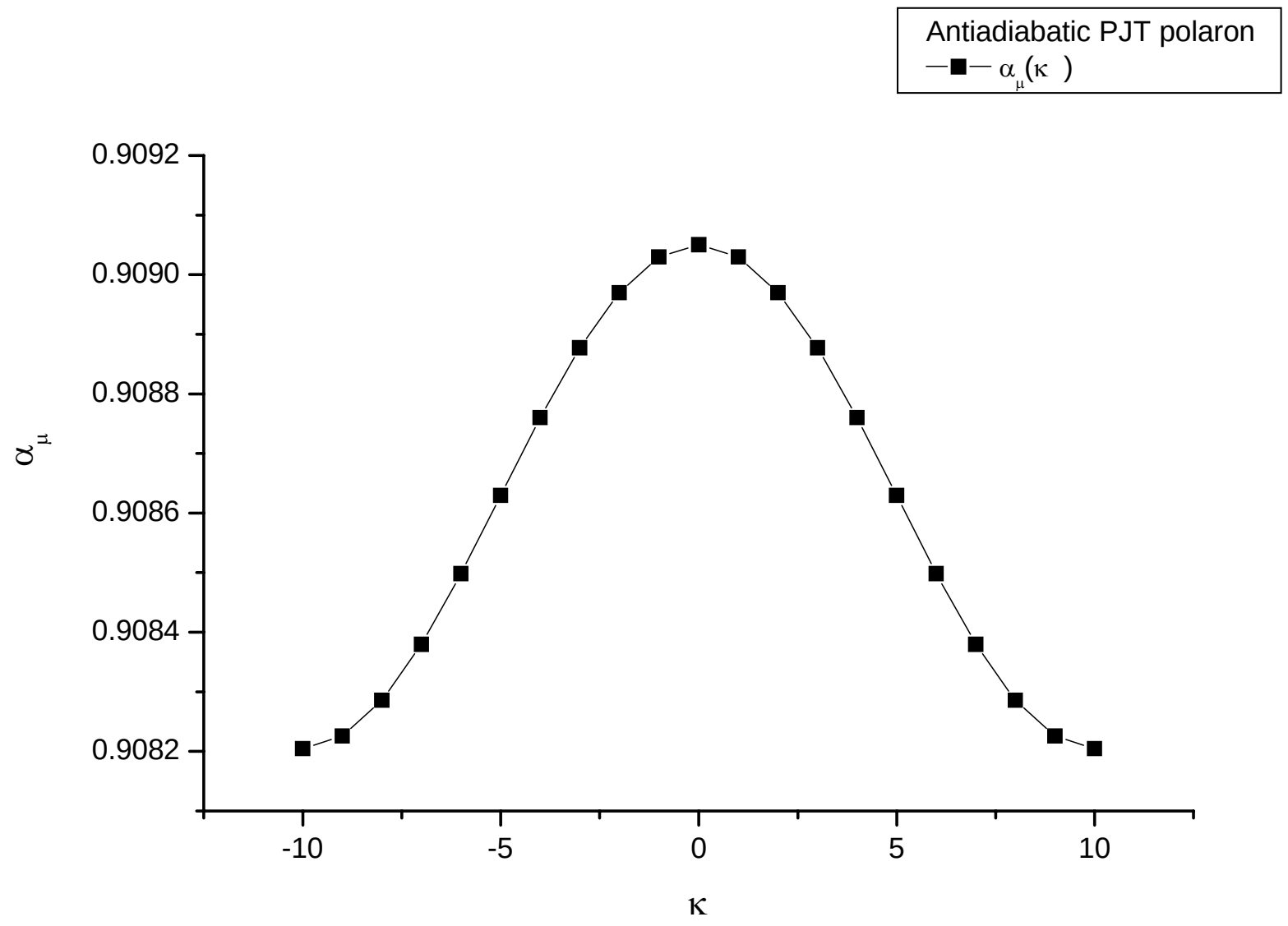

Figure 4 (g)

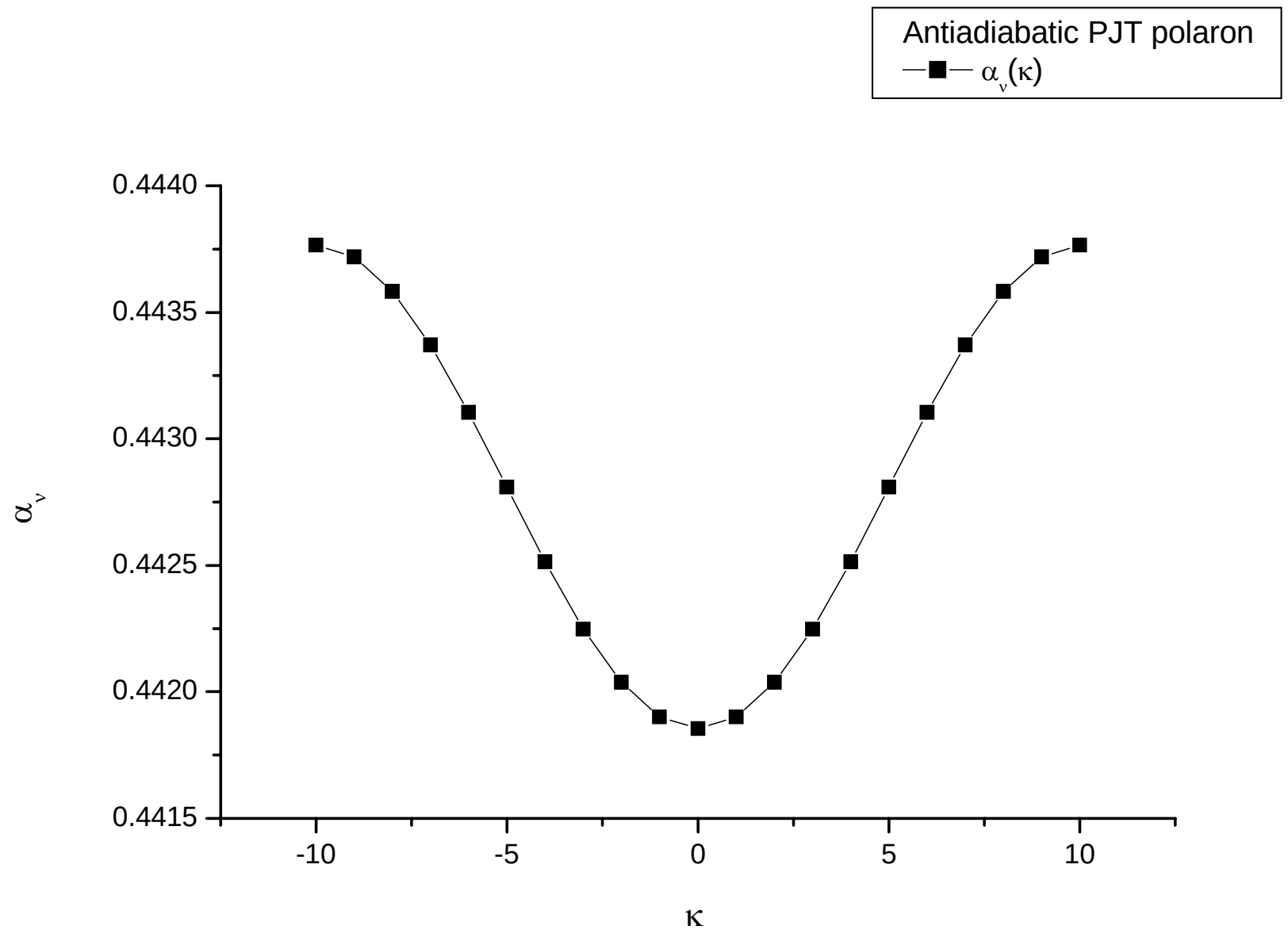

Figure 4 (h)

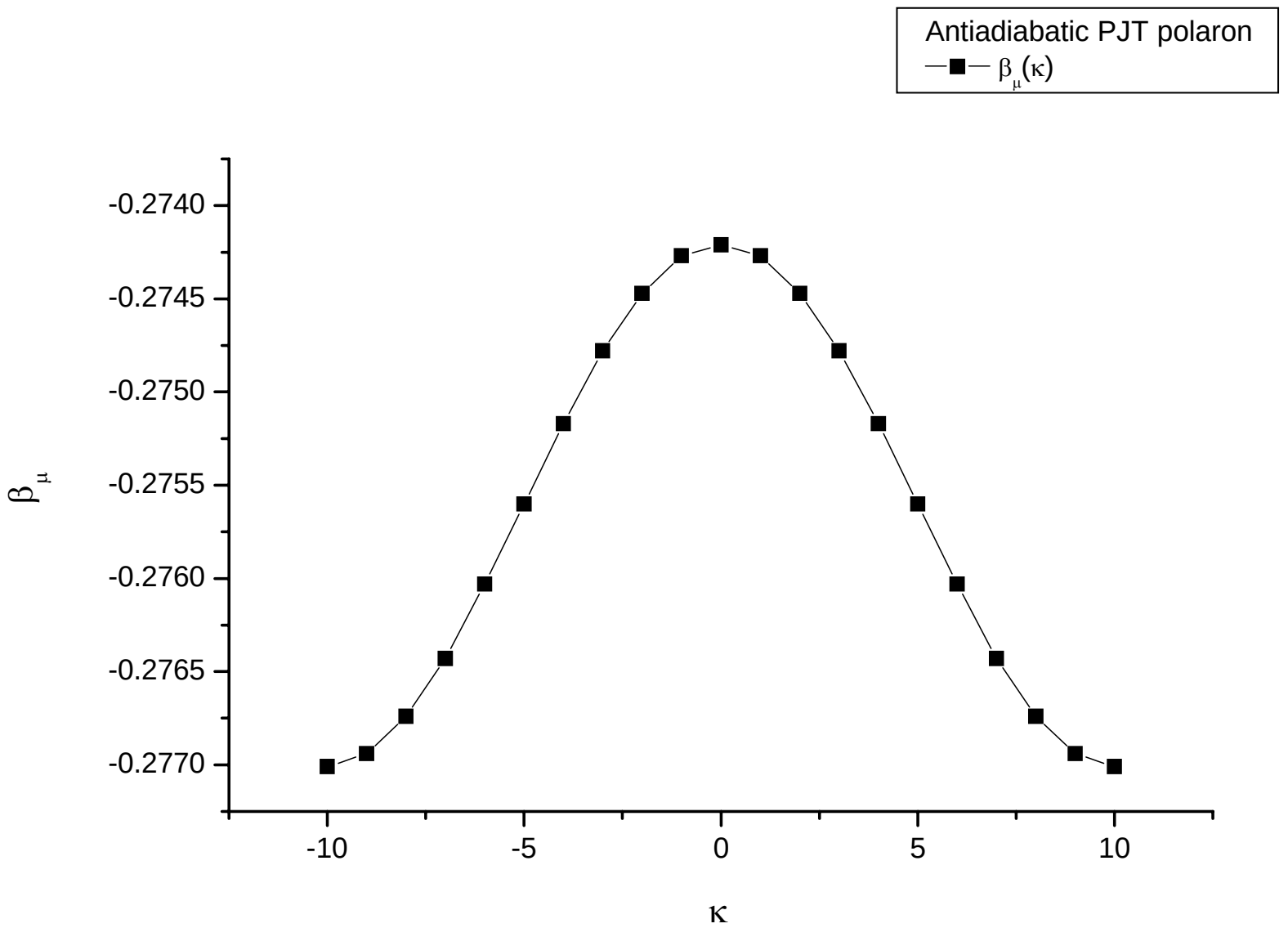

Figure 4 (i)

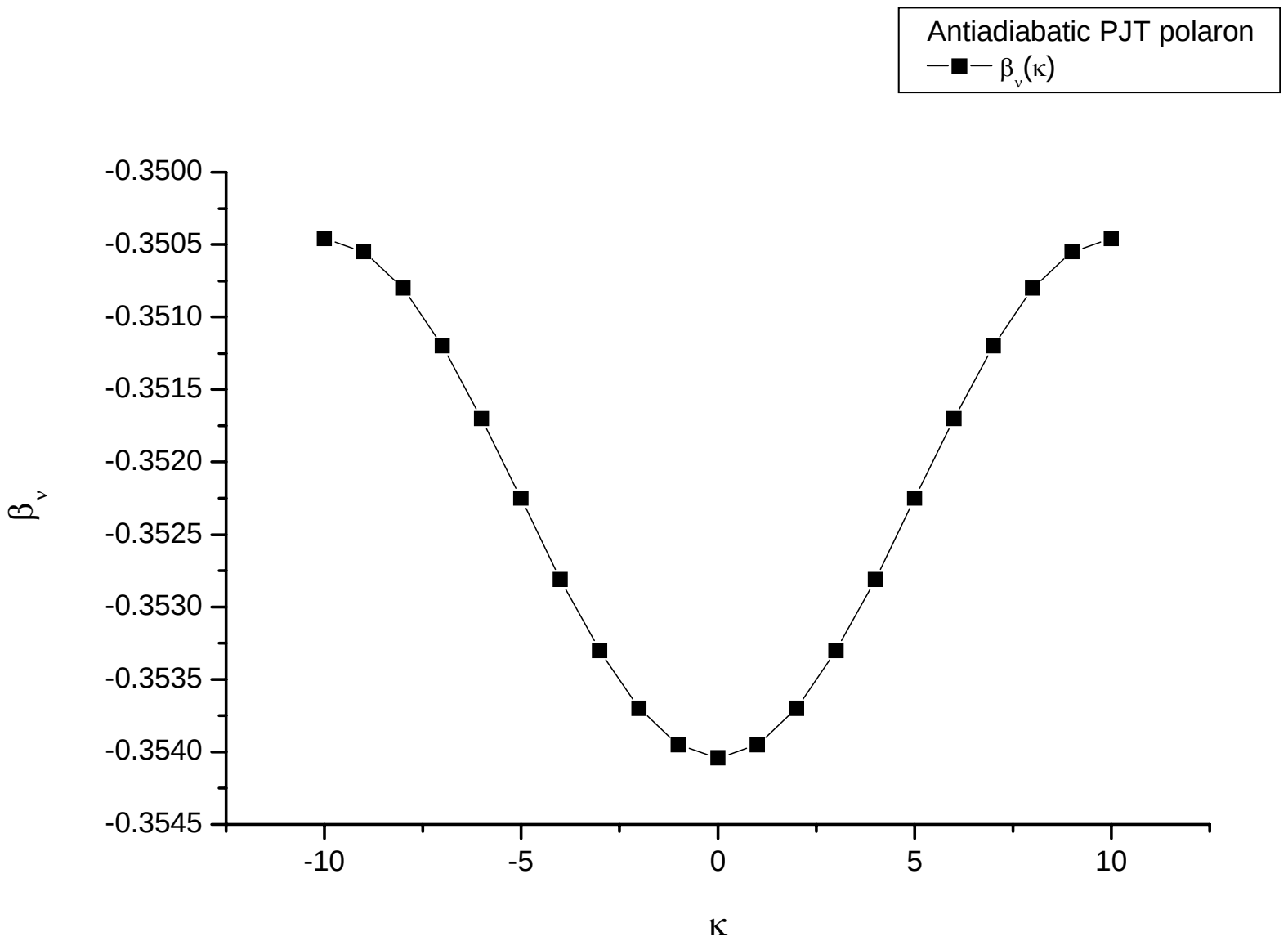

Figure 4 (j)

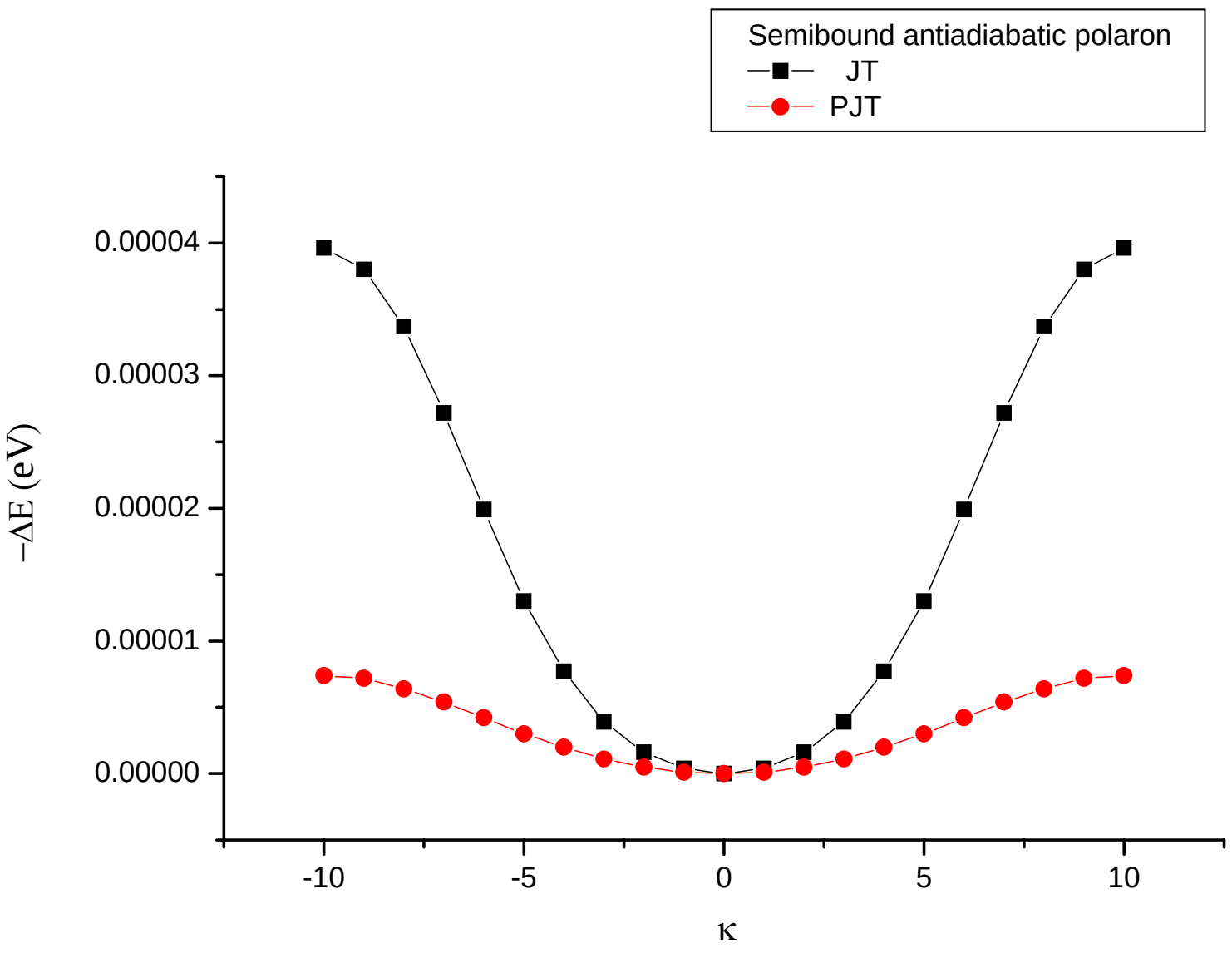

Figure 5

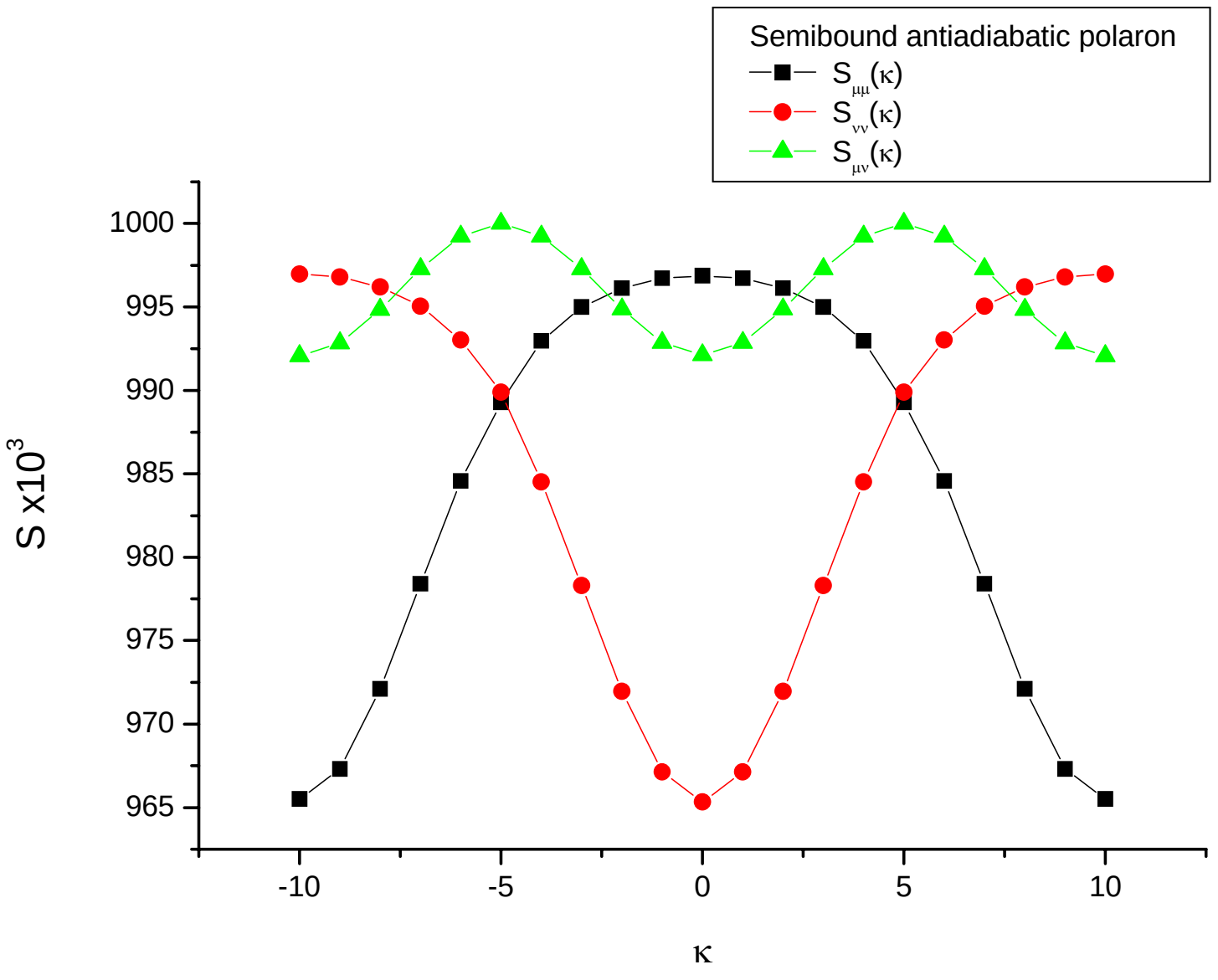

Figure 6 (a)

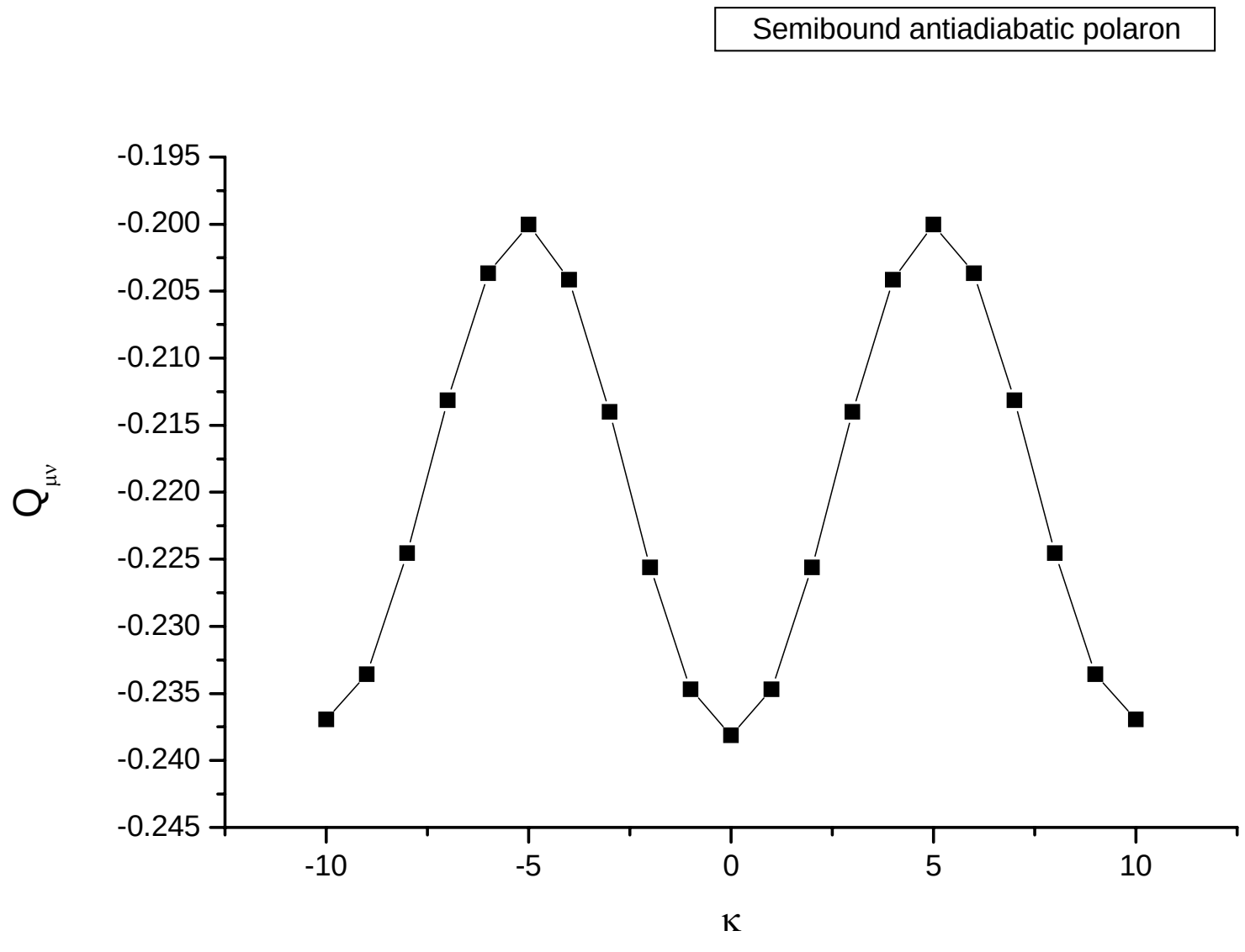

Figure 6 (b)

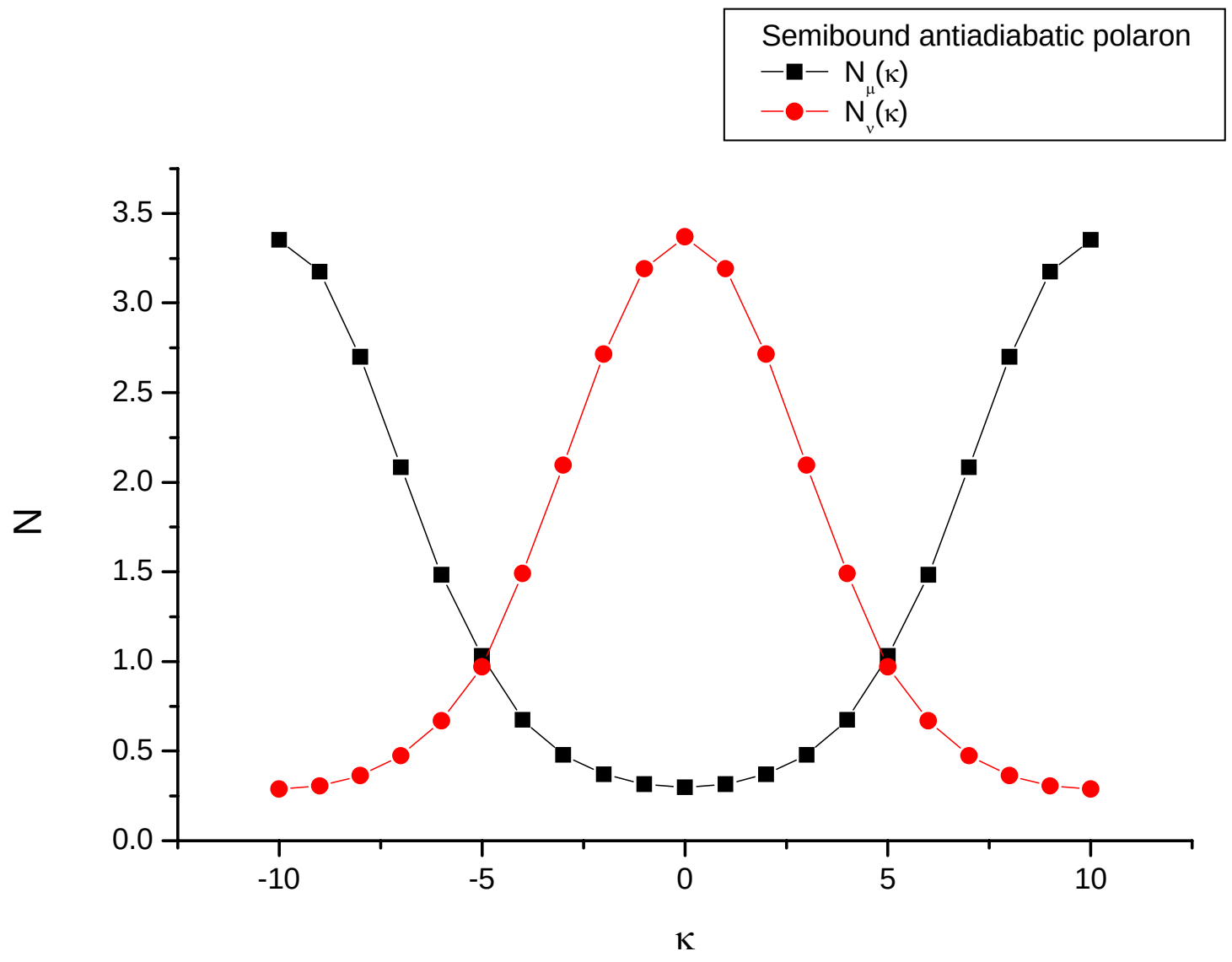

Figure 6 ( c)

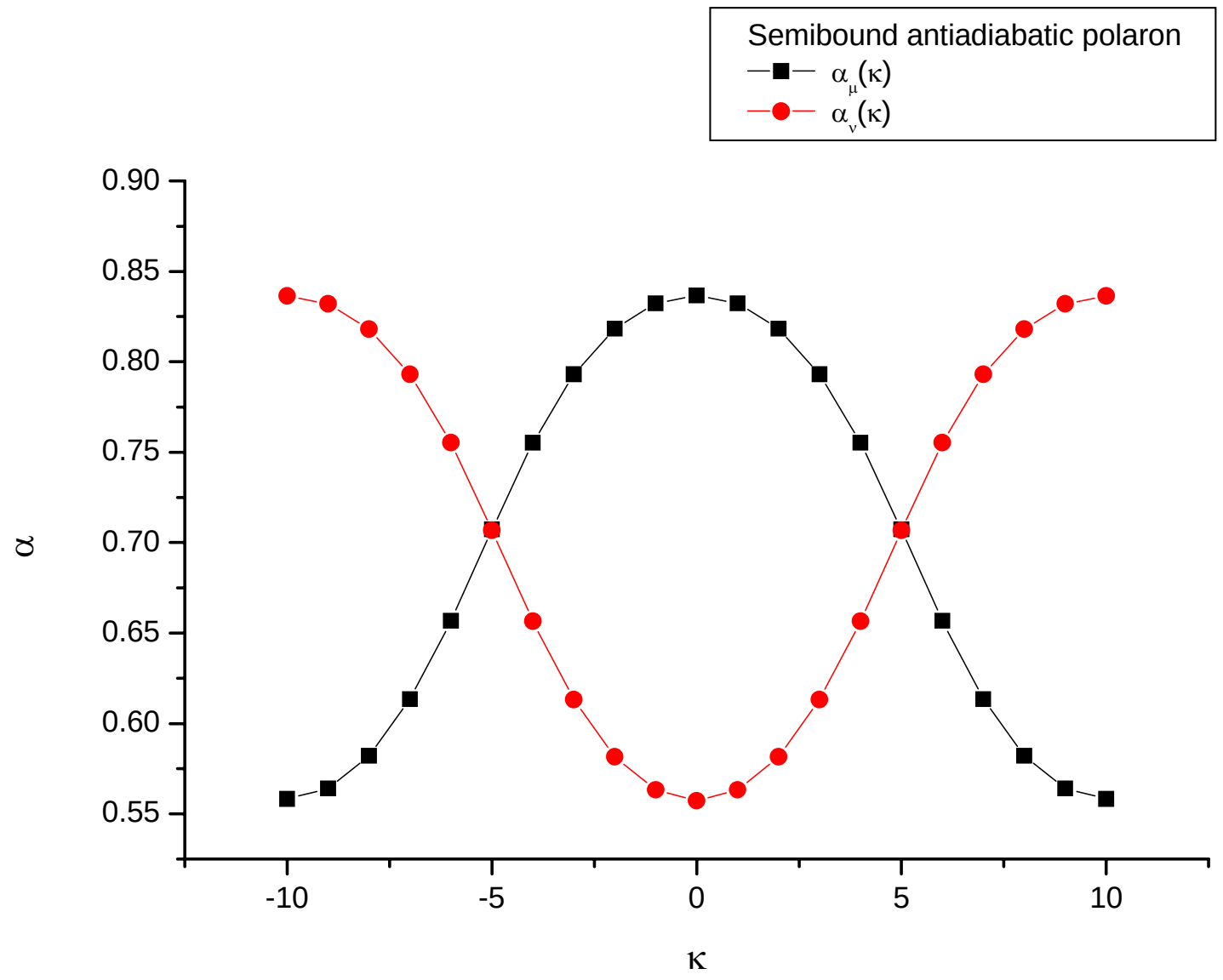

Figure 6 (d)

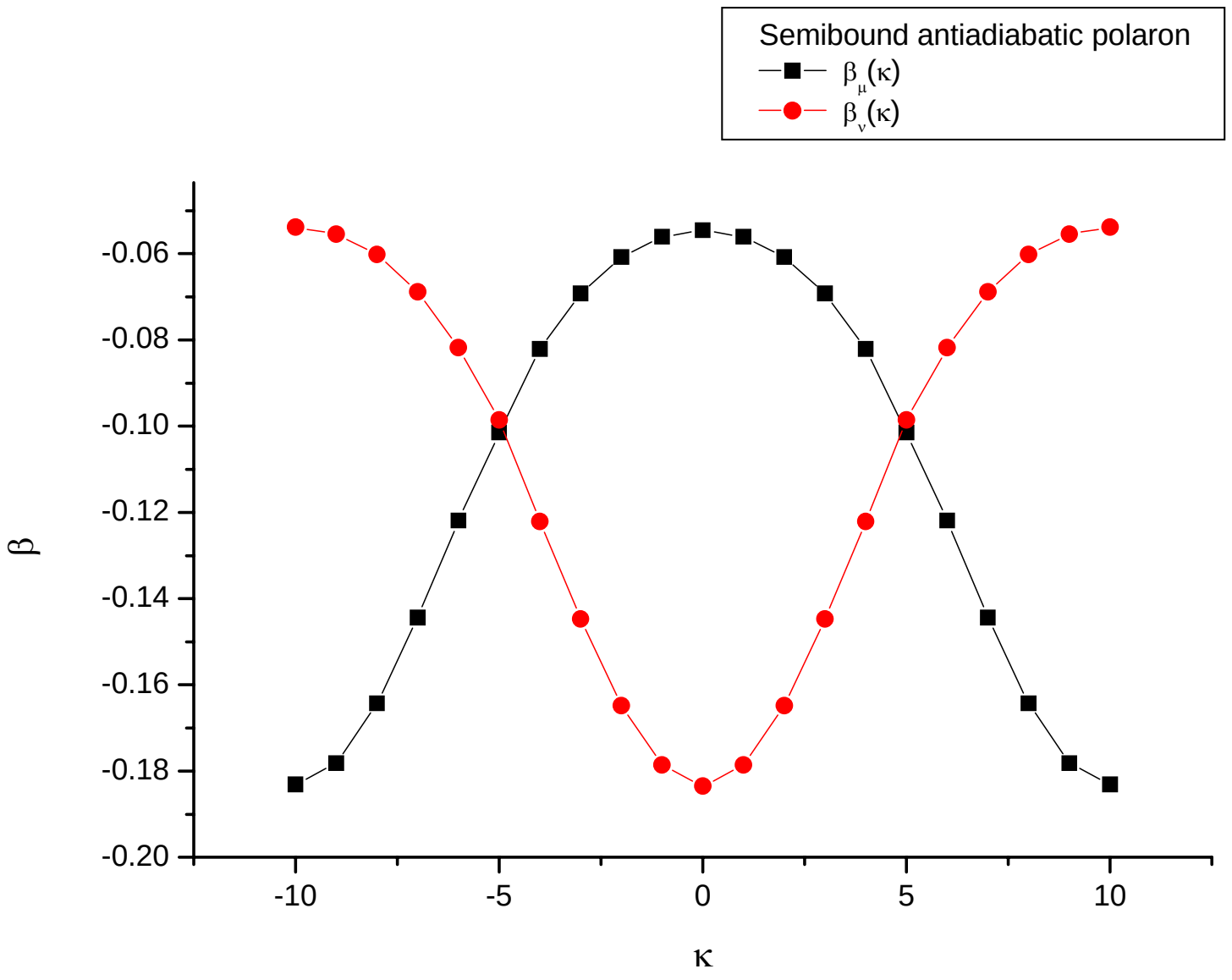

Figure 6 (e)

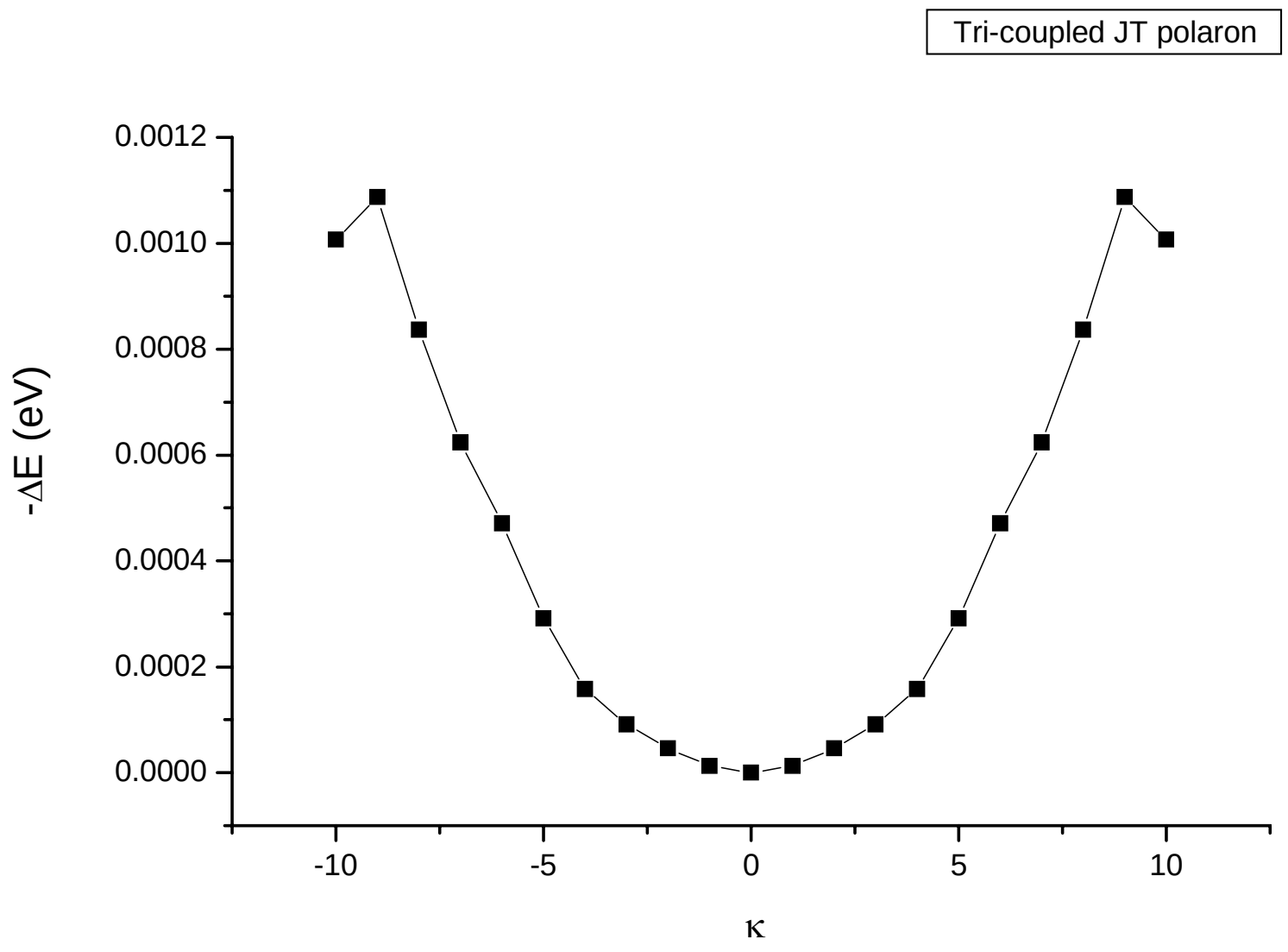

Figure 7